\documentclass{article}
\usepackage{graphicx} 
\usepackage[left]{lineno} 
\usepackage[a4paper, margin=1in]{geometry} 
\usepackage{pdflscape}
\usepackage{breqn}
\usepackage{breqn}
\usepackage{xcolor}
\usepackage{amssymb}     
\usepackage{amsfonts}    
\usepackage{mathtools}   
\usepackage{bm}          
\usepackage{algorithm}        
\usepackage{algorithmic}      
\usepackage{listings}         
\usepackage{booktabs}    
\usepackage{array}       
\usepackage{multirow}    
\usepackage{tabularx}    
\usepackage{textcomp}
\usepackage{capt-of}
\usepackage{array}
\usepackage{booktabs}
\usepackage{placeins}
\usepackage{needspace}
\newcolumntype{L}[1]{>{\raggedright\arraybackslash}p{#1}}

\usepackage{booktabs,tabularx,array}
\usepackage{setspace}
\usepackage{graphicx} 
\usepackage[left]{lineno} 
\usepackage[a4paper, margin=1in]{geometry} 
\usepackage{amsmath}     
\usepackage{amssymb}     
\usepackage{amsfonts}    
\usepackage{mathtools}   
\usepackage{bm}          
\usepackage{algorithm}        
\usepackage{algorithmic}      
\usepackage{listings}         
\usepackage{booktabs}    
\usepackage{array}       
\usepackage{multirow}    
\usepackage{tabularx}    
\usepackage{setspace}    

\usepackage[style=nature, sorting=none, backend=bibtex]{biblatex}
\date{} 

\begin{document}



\begin{center}

  {\LARGE\bfseries
  ADF-EA: A Unified Execution Assurance System
  for Agent Device Foundation\par}
  \vspace{1em}
  {\large
  Xuechun Li,
  Jiaxin Liang,
  Jie Li,
  Baolong Li, 
  Jue Wang,
  Peng Yuan,
  Hang Huang}
  \vspace{0.5em}
  {\normalsize\\
  Huawei\par}

  \vspace{0.3em}
\end{center}
\vspace{0.3cm}


\begin{abstract}
Agents based on large language models (LLMs) can access heterogeneous devices through tools and APIs, enabling increasingly complex tasks across devices. As agents gain greater autonomy in planning and recovery, safe and reliable execution must account for unmet effects, uncertain outcomes, and changing prerequisites. A command may be acknowledged without producing its intended effect, while missing feedback may obscure an action that has already succeeded. Across heterogeneous devices, capability knowledge must support both flexible planning and consistent execution decisions. We present Agent Device Foundation--Execution Assurance (ADF-EA), an architecture connecting agent planning and device execution through shared capability contracts. Device Capability Contracts (DCCs) unify invocation conditions, intended effects, evidence requirements, and recovery rules across heterogeneous interfaces. Agents use these contracts to plan, while the runtime applies the same semantics to authorize actions, verify effects, and govern continuation and completion. Persistent execution state retains verified progress, unresolved outcomes, and remaining budgets across plan revisions, supporting observation-based recovery, authorized retries, and necessary state repair. We formalize the execution lifecycle and establish conditional soundness properties for completion and recovery authorization. Evaluations span multiple LLMs, five agent frameworks, and simulated process-control, household, and robotic manipulation domains. Comparisons with direct invocation and existing execution-checking approaches demonstrate gains in execution assurance while preserving permitted task completion and recovery. ADF-EA reduces false completion and unnecessary repetition, supports necessary state repair, and prevents calls to unavailable capabilities. These results demonstrate DCCs as a reusable semantic foundation for agent autonomy across heterogeneous devices, unifying capability-based planning, evidence-grounded execution, and authorized recovery within one architecture.
\end{abstract}

\section{Introduction}
\label{sec:introduction}

Agents based on large language models (LLMs) can access devices through tools and APIs, creating opportunities to automate tasks across household appliances, laboratory instruments, and robotic systems. As agents take on greater autonomy in selecting actions, coordinating devices, and recovering from failures, ensuring safe and reliable execution becomes a central challenge. Their decisions must respect current device conditions, account for the effects of earlier actions, and establish that the intended task outcomes have been achieved.

Physical-device execution makes these requirements consequential. A command may be acknowledged without producing its intended effect, a timeout may follow an action that has already succeeded, and a previously verified state may change before a dependent action begins. In a multi-device task, continuing without an established prerequisite can invalidate later actions; repeating an action with an uncertain outcome can duplicate physical work. Reliable recovery therefore requires distinguishing effect failure, insufficient evidence, and invalidated prerequisites, retaining valid progress, and determining which further actions are permitted. These execution questions remain even when every device interface is accessible and every request is syntactically valid.

Applying these rules across heterogeneous devices also requires a reusable representation of capability knowledge. Vendors expose different interfaces, while operating requirements are distributed across device models, manuals, procedures, and tool descriptions. Tool descriptions can already express capabilities, parameters, and constraints. A shared contract can organize this information into invocation conditions, intended effects, evidence requirements, and recovery permissions that guide both planning and runtime decisions. Such a representation provides a basis for applying the same execution-assurance rules across devices and invocation mechanisms.

Feedback-driven agents can revise their next action \cite{yao2022react,huang2022inner}, while the effects and unresolved outcomes of earlier actions continue to constrain what may happen next. Our central question is: \emph{How can agents retain flexible planning and recovery while device execution remains grounded in verified progress and governed by shared capability rules?}

We present \textbf{Agent Device Foundation--Execution Assurance (ADF-EA)}, a contract-directed architecture connecting heterogeneous device knowledge, LLM planning, and execution assurance. \textbf{Device Capability Contracts (DCCs)} organize invocation conditions, intended effects, evidence requirements, and interruption/recovery rules into a shared semantic layer above Model Context Protocol (MCP) tools, command-line interfaces (CLIs), software development kits (SDKs), and other invocation mechanisms. The agent uses these contracts to select and compose capabilities; the runtime applies the same contracts to authorize actions, establish verified progress, and govern continuation and completion. This connects what the agent understands a capability to mean with the rules that govern its execution.

Within this architecture, revisable proposals are separated from persistent execution state. An uncertain outcome remains attached to its operation, whereas a historical success is distinguished from the current validity of the state it established. Consider placing an object in a cabinet and leaving the cabinet closed. If a verified open cabinet subsequently closes, reopening repairs the prerequisite for placement. If placement executes but its evidence is missing, another placement request instead concerns an unresolved outcome. The first case calls for necessary state repair; the second calls for evidence acquisition or stopping, unless qualified negative evidence and the contract authorize a retry. The same architecture thus supports recovery without treating uncertainty as permission for repeated physical work.

The architectural contribution lies in how these elements govern a continuing task. Evidence changes usable progress; usable progress and unresolved obligations constrain the next authorization; residual-plan revision preserves that state and its remaining budgets. Individual checks are thus connected by persistent execution semantics rather than evaluated as unrelated calls. The formal model characterizes the information boundary and establishes the completion and recovery properties preserved by these transitions.

Evaluation across simulated process-control, household, and robotic manipulation tasks demonstrates complementary gains in recovery, execution control, and reuse. The primary controlled comparison evaluates ADF-EA against public-algorithm reproductions of ToolGate and VTC and a direct-invocation baseline. ADF-EA completes all 14 completion-designated conditions with qualified evidence and no violating dispatches. Autonomous household case studies also complete all three required state repairs and record no improper repetitions under persistent evidence loss. Across five agent frameworks, it preserves all fifty permitted completions while blocking all fifty unavailable-provider calls made by Direct. Studies with multiple LLMs assess both autonomous decision-making and execution of model-generated proposals, showing consistent reductions in false completion and disallowed dispatches. Controlled comparisons and ablations identify how contract semantics, evidence qualification, and persistent execution state produce these benefits.

The paper makes three contributions:
\begin{itemize}
\item \textbf{Persistent execution semantics for adaptive agents.} We organize execution uncertainty, invalidated prerequisites, and renewed authority in a state model that survives requests and residual-plan revisions. Distinct retry and state-repair rules preserve unresolved obligations, verified history, and remaining budgets.
\item \textbf{A reusable contract-directed agent architecture.} DCC connects heterogeneous capability knowledge to planner-facing descriptions and execution-facing checks above concrete invocation interfaces. Domain contracts and adapters instantiate the shared lifecycle for device actions and high-level skills.
\item \textbf{Formal properties and demonstrated behavioral benefits.} Two conditional soundness results establish completion and recovery-authorization properties of the transition system. Autonomous comparisons preserve task completion and recovery while eliminating unsupported repetition and disallowed dispatches. Controlled studies identify the contributing mechanisms, and evaluations across models, agent frameworks, and device domains demonstrate architectural reuse.
\end{itemize}

The method develops the capability representation and its runtime state, then states the authority transitions and their properties. The evaluation examines recovery utility, execution control, semantic attribution, and reuse. Detailed implementation and supporting experimental records are retained in the appendices.

\section{Related Work}
\label{sec:related-work}

\subsection{Adaptive Planning and Execution Authority}

Foundation-model systems expand how agents select and compose physical actions. SayCan grounds language proposals in skill affordances, while Code as Policies produces programs over perception and control APIs. PaLM-E and RT-2 connect perception and language to robot actions. VoxPoser uses language to construct spatial constraints \cite{ahn2022can,liang2023code,driess2023palm,brohan2023rt,huang2023voxposer}. Inner Monologue and ReAct use feedback to revise subsequent behavior \cite{huang2022inner,yao2022react}. Replanning changes the proposed future, but does not establish the physical outcome of past actions. ADF-EA therefore separates proposal state from authoritative execution state, using qualified effect evidence to determine progress and renewed device authorization.

Behavior Trees and robot executives already provide sequencing, monitoring, and recovery \cite{ogren2018behavior}. Simplex separates advanced controllers from trusted safety mechanisms, and SOTER supports reactive programming and composition of runtime-assurance modules \cite{sha2001using,phan2017component,mehmood2022black,desai2019soter}. These establish foundations for governing execution independently of the proposing controller. ADF-EA applies that separation to capability-level task state: evidence of an effect determines which progress is accepted and what continuation remains admissible within the represented task and provider configuration.

\subsection{Safety Contracts, Typed Skills, and Execution Governance}

ToolGate represents trusted world information in typed symbolic state and uses Hoare-style preconditions and postconditions to gate invocation and verify state updates \cite{liu2026toolgate}. Verified Tool Calls addresses ambiguous outcomes through postcondition verification, bounded verify-before-retry, and idempotency keys \cite{mansoor2026verifiedtoolcalls}. These approaches illustrate how contracts and verification govern tool execution. ADF-EA uses persistent task obligations and operation-scoped authority to connect verification with subsequent execution: an unresolved outcome constrains later requests, while a verified state that subsequently becomes false can authorize necessary repair. This distinction retains execution meaning across calls and plan revisions.

Safety Chip, SELP, and VerifyLLM constrain model-generated behavior using explicit specifications \cite{yang2024plug,wu2025selp,grigorev2025verifyllm}. SafeAgentBench supplies ThinkSafe, a next-action safety checker, while RoboGuard places enforcement downstream of the planner \cite{yin2024safeagentbench,ravichandran2026safety}. Safety-contract and execution-governance systems address additional parts of the lifecycle. SafeGate combines guards, invariants, abort conditions, runtime pre/post-action checks, state tracking, and violation records \cite{obi2026pre}. ROSClaw by Cardenas et al. describes capability discovery, action validation, normalized observations, and audit logging \cite{cardenas2026rosclaw}. Separately, the \texttt{ros-claw/rosclaw} runtime repository documents expected-effect verification, evidence-bearing receipts, and durable action state \cite{rosclaw_runtime_2026}. Its documented restart handling preserves unknown physical outcomes for interrupted REAL actions, prevents automatic re-execution, requests an emergency stop, and requires operator review before lifting the recovery gate. We assess these sources separately rather than treating the repository as the implementation accompanying the Cardenas et al. paper.

Typed-skill orchestration is another close comparison. Physical Agentic AI exposes typed parameters, preconditions, expected effects, and workflow completion conditions, with deterministic per-step authorization and mission termination after execution failure \cite{liu2026physicalagenticaiarchitecture}. SINT documents capability authorization, request states, failure/rollback transitions, and a hash-linked evidence ledger \cite{sint_protocol_repo_2026}. Table~\ref{tab:execution-assurance-comparison} summarizes the overlap using specific representatives of the broader families and identifying the execution decisions supported by their documented mechanisms. Capability interfaces are discussed separately in Section~\ref{sec:capability-interfaces}.

\begin{table*}[t]
\centering
\small
\setlength{\tabcolsep}{4pt}
\renewcommand{\arraystretch}{1.12}
\caption{Positioning ADF-EA relative to representative execution-governance systems.
Entries summarize documented contracts, execution state, recovery mechanisms,
and the decisions they govern. Device-access interfaces are discussed separately.}
\label{tab:execution-assurance-comparison}
\par\vspace{6pt}

\begin{tabularx}{\textwidth}{@{}>{\raggedright\arraybackslash}p{0.16\textwidth}*{3}{>{\raggedright\arraybackslash}X}@{}}
\toprule
System & Contracts and checks & Execution state and recovery & Assurance focus \\
\midrule
ToolGate~\cite{liu2026toolgate}
& Hoare-style preconditions and postconditions
& Typed symbolic state; state updates committed after verification
& Admissible invocation and verified state evolution \\
\addlinespace
ROSClaw (Cardenas et al.)~\cite{cardenas2026rosclaw}
& Capability discovery; pre-execution action validation
& Normalized observations; structured audit logs
& Mediated ROS~2 execution and behavioral auditing \\
\addlinespace
ROSClaw runtime repository~\cite{rosclaw_runtime_2026}
& Action authorization; effect verification
& Durable receipts; unknown interrupted REAL outcomes; no automatic retry; operator review
& Evidence-backed execution records and restart gating \\
\addlinespace
SafeGate~\cite{obi2026pre}
& Guards, invariants, aborts; pre/post-action checks
& Symbolic state; violation records; halt on violation
& Safety-contract compliance during execution \\
\addlinespace
SINT~\cite{sint_protocol_repo_2026}
& Capability tokens; policy and physical constraints
& Request lifecycle; evidence ledger; failure/rollback transitions
& Authorized requests and auditable execution records \\
\addlinespace
RTA: SOTER~\cite{desai2019soter}
& Safety specifications; runtime monitoring
& Controller state; safe-controller fallback
& Safety preservation under module assumptions \\
\addlinespace
Typed skills: Physical Agentic AI~\cite{liu2026physicalagenticaiarchitecture}
& Typed parameters; expected effects; per-step authorization
& Robot/workflow state; structured feedback; halt on failure
& Validated orchestration of robot skills \\
\addlinespace
ADF-EA
& Invocation, effect, evidence, and interruption conditions
& Persistent progress and obligations; observation repair; distinct retry and state repair; permitted rebinding
& Evidence-grounded authority across requests, plan revisions, and completion \\
\bottomrule
\end{tabularx}
\par\smallskip
\parbox{\textwidth}{\footnotesize The ROSClaw paper and the \texttt{ros-claw/rosclaw} runtime repository are assessed separately; no implementation lineage between them is assumed. The runtime repository and SINT entries describe documented mechanisms, not independently validated guarantees. Repository access dates are recorded in the corresponding references.}
\end{table*}

ADF-EA makes effect evidence and execution history a common basis for task progress, renewed authority, and completion. Its persistent lifecycle connects the result of one check to the permissions governing subsequent requests. Qualified evidence of an absent effect can support a permitted retry; insufficient evidence retains the obligation and calls for observation; a previously established prerequisite that becomes false can require state repair. These distinctions remain in force when an agent revises its proposal. DCC makes the conditions explicit across capability interfaces, and execution assurance applies them throughout the task.

\subsection{Capability Interfaces and Semantic Reuse}
\label{sec:capability-interfaces}

Device-access standards provide reusable descriptions and interaction mechanisms for heterogeneous devices. W3C Web of Things (WoT) Thing Description 1.1 describes properties, actions, and events together with data schemas, security metadata, and protocol bindings. It also represents action idempotence and a \texttt{safe} flag describing invocations that do not change internal resource state \cite{w3c2023wottd11}. The Model Context Protocol (MCP) specifies tool discovery and invocation, input and optional output schemas, and structured results. Its tools specification distinguishes behavioral annotations from trusted information by requiring clients to treat annotations from untrusted servers as untrusted \cite{mcp2026tools}.

Recent device-agent systems bring these interfaces into physical workflows. Arm's Device Connect supports device discovery, publish/subscribe state exchange, and remote procedure calls through a shared driver model \cite{arm2026deviceconnect}. The Model Hardware Standard (MHS) research preview describes standardized device drivers with read/write interfaces, operational metadata, and enforced device-level safety limits. Its reported applications include monitoring, adaptive orchestration, and recovery from hardware errors \cite{anthropic2026mhs}. These systems already support aspects of constrained, feedback-driven device operation.

ADF-EA positions DCC above concrete invocation interfaces as a shared capability representation for planning and execution. Its assurance lifecycle determines how observations and execution history acquire authority over subsequent task execution. A callable operation, a structured result, or an observable property supplies information that a DCC relates to a specific capability invocation. Its evidence semantics determine whether that information establishes the required effect. Its interruption semantics determine which established effects remain usable after disruption. The resulting verified progress and unresolved obligations govern whether execution may continue, repeat physical work, use a permitted replacement provider, or declare completion. This evidence-to-authority lifecycle provides the assurance semantics that an access interface can carry through domain-specific adapters. The distinction concerns the decisions made from device information, including when further observation is required before physical recovery can be authorized.

\section{ADF-EA: Agent Architecture and Persistent Execution Semantics}
\label{sec:adf-ea}

\begin{figure}[t]
\centering
\includegraphics[width=\linewidth]{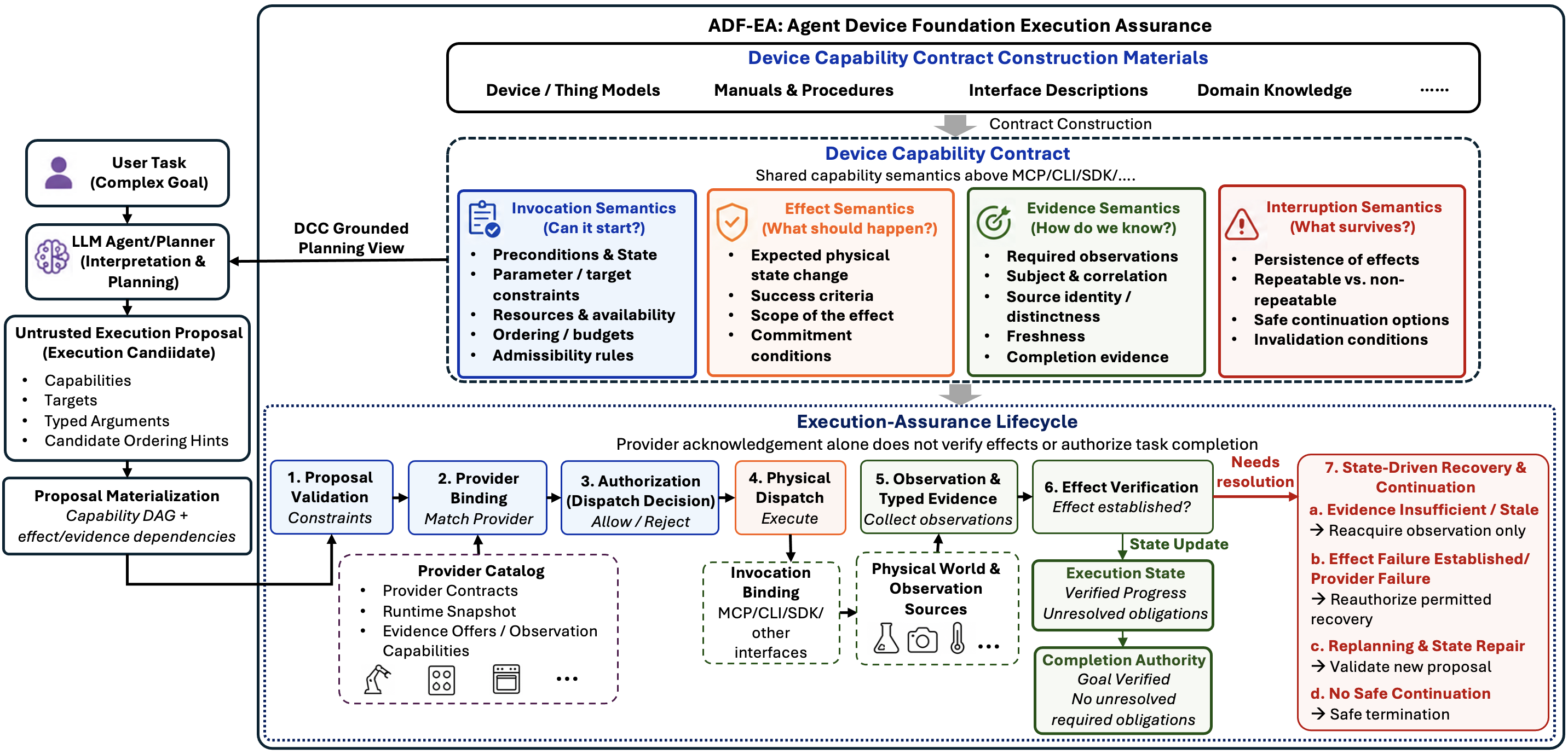}
\caption{Overview of ADF-EA. Heterogeneous device knowledge is organized into Device Capability Contracts (DCCs), which provide shared semantics for LLM planning and execution assurance above MCP, CLI, SDK, and other invocation interfaces. Contract-directed authorization and evidence qualification govern dispatch, task progress, recovery, and completion. Recovery includes observation reacquisition, permitted retries, and necessary state repair through validated plan revisions.}
\label{fig:adf-ea-overview}
\end{figure}

Figure~\ref{fig:adf-ea-overview} shows how a revisable agent proposal interacts with persistent execution state. DCC supplies capability meaning to both planning and enforcement; the lifecycle retains the history and obligations needed to authorize continued execution. We develop the architecture and contract representation, then formalize its transitions and the properties they preserve.

\subsection{DCC-Grounded Planning and Execution Authority}
\label{sec:agent-authority}

The system connects a DCC-grounded capability catalog, an upstream LLM agent, execution assurance, device providers, and observation sources. The agent uses capability descriptions, represented dependencies, the task, and current public feedback to select and compose actions. It submits an inspectable candidate containing capability identifiers, arguments, targets, and proposed ordering. Providers realize the selected capabilities through instrument commands, household actions, or bounded robotic skills; observation sources supply measurements or structured state. EA interprets the candidate against the corresponding contract and task semantics, mediates dispatch, and returns execution feedback for the agent's next proposal.

Execution state distinguishes an attempted invocation, evidence supporting its effect, and progress accepted for the task. A tool response records an interaction outcome. It does not independently establish accepted progress. For example, a successful Put response can be recorded while placement remains unverified. A dependent Close can proceed only when its represented prerequisites have been established. In a sequential task, accepted progress may form a verified prefix. In a DAG, it consists of established dependencies and remaining obligations.

The authority model assumes a declared task specification and mediation of all relevant device actions. Contracts and observation adapters supply the physical meaning of its decisions. Progress retention and recovery operate over the task dependencies and interruption rules represented in the runtime. A proposed change is subject to validation and does not by itself certify a previous effect or grant permission for a new invocation.

\label{sec:running-example}
The task is to pick up an object, open a cabinet, place the object inside, and close the cabinet. Its execution history determines what a repeated request means:
\begin{enumerate}
\item If an Open attempt has an unresolved outcome and qualified negative readback establishes failure, a contract-permitted retry may restore progress.
\item If Open was verified but the cabinet later closes, reopening is a necessary repair of the state required by Put, not failure of the earlier Open.
\item If Put executes but its evidence remains unknown, the obligation stays open. A further Put request cannot bypass it; allowed observation may establish placement and enable Close.
\item If a relevant state-setting effect still holds, repeating the action without a new authorized purpose is redundant and is rejected.
\end{enumerate}
These decisions use the same retained history and current evidence. State repair preserves the ledger and remaining budget; only new qualified evidence establishes new progress. Sections~\ref{sec:authority-transitions} and~\ref{sec:conditional-soundness} formalize these rules, and Section~\ref{sec:native-agent-mechanism-comparison} tests the repair and unknown-evidence branches with autonomous agents.

\subsection{Device Capability Contracts}
\label{sec:capability-contracts}

A Device Capability Contract is the common semantic unit for capability planning and execution assurance. For capability \(c\), write \(\mathcal C_c=(\mathcal I_c,\mathcal E_c,\mathcal V_c,\mathcal R_c)\). Invocation conditions \(\mathcal I_c\) describe admissible use. Intended effects \(\mathcal E_c\) describe what the planner can seek to establish and connect to downstream prerequisites. Evidence requirements \(\mathcal V_c\) determine which observations justify treating those effects as achieved. Interruption rules \(\mathcal R_c\) govern observation, retry, and continued execution when progress is disrupted. Together, the four blocks connect planned capability use to the evidence and authority required for its realization.

\label{sec:dcc-construction}
DCC construction draws on an open-ended set of relevant knowledge, including device or thing models, equipment manuals, operating procedures, API documentation, MCP tool descriptions, CLI help, SDK declarations, and domain-authored specifications. Capability information is mapped to the four semantic blocks and completed with domain knowledge. These construction materials are distinct from runtime invocation arguments.

Device models and operating descriptions can supply state predicates and effects; interface declarations supply callable identities and typed arguments; observation specifications supply evidence requirements; and operating policies supply recovery rules. No single source must provide every block. Automated extraction and manual authoring can feed this boundary; the evaluated contracts use authored declarations and domain projections into normalized semantics.

\label{sec:dcc-interface-layer}
Construction provenance and execution binding are independent: a manual can specify an effect of a capability invoked through an SDK. Provider bindings resolve capabilities to invocation and observation routes through MCP, CLI, SDK, or other interfaces; DCC supplies their capability semantics without replacing transport or calling conventions.

Domain adapters also determine action granularity: a household command establishes an object state, an instrument operation a measured process condition, and a manipulation skill a goal through multiple controller steps. Each supplies its own effect and evidence predicates to the shared contract structure.

\label{sec:dcc-planning-view}
The planner-facing view exposes capability identity, targets, input constraints, preconditions, effects, and task-relevant evidence and recovery requirements. The LLM uses it with the task and public state to select capabilities and propose ordering. An intended effect can support a planned dependency; only qualified evidence makes that dependency usable execution progress.

The execution-facing view adds provider bindings and observation routes for admission, qualification, and progress updates. Feedback returns these outcomes to the planner without exposing internal contract or ledger objects. In the native Agent path, public action descriptions and prerequisite metadata form the planning view, while the generic projection supplies runtime semantics. The autonomous native case studies use this public action interface and its declared capability projection.

\label{sec:dcc-worked-example}
In the evaluated four-step cabinet task, the action declarations contain typed metadata named \texttt{requires} and \texttt{declared\_effects}. Open, the second step, requires \texttt{sequence.ready.02}; its declared effects include establishing \texttt{sequence.ready.03}, which Put requires. These declarations and grounded object identifiers supply the capability projection. The readiness predicate is interpreted using verified predecessor effects and the currently observed open state of the target receptacle. It is therefore a task dependency with physical meaning, not just an index in a fixed script.

The four DCC blocks connect this declaration to execution. \(\mathcal I_{\mathrm{Open}}\) captures invocation readiness and target constraints; \(\mathcal E_{\mathrm{Open}}\) establishes the state needed for placement; \(\mathcal V_{\mathrm{Open}}\) associates the effect with qualified, object-specific readback; and \(\mathcal R_{\mathrm{Open}}\) governs admissible recovery when execution or its evidence is disrupted. The native Agent sees action descriptions and dependency metadata, with scripted ordering hints removed, and chooses its own next request. The runtime uses the corresponding normalized semantics to decide which of those requests may execute.

If a later observation shows the cabinet closed, the adapter withdraws Put readiness but retains the verified Open history, supporting state-repair admission. If Put executes without qualifying evidence, readiness for the fourth step remains false and its execution obligation stays open. These projected states connect the concrete declarations to the transition rules below.

The public DCC format requires a specification version, capability identity, subject, and all four semantic blocks. Effects have explicit identifiers. Evidence requirements reference them and specify source, freshness, and observation-count conditions. Predicates contain a subject and state path, typed expected value, and supported comparison operator. Deployable-contract validation checks structure, unique identifiers, evidence coverage, references, and interruption-cause coverage. Provider acknowledgements are retained only as invocation-lineage records. They do not establish physical effects or task completion. The implementation predates the paper's ADF naming and therefore retains the artifact identifier \texttt{ADA-DCC v0.1}. The paper uses DCC for the capability-contract abstraction.

Provider-bound runtime contracts add provider and tool identities, input constraints, a stable contract reference, semantic version, lifecycle state, and evidence routes. Their fingerprints identify the contract content captured for an execution. The runtime also accepts domain projections into normalized capability semantics. Heating uses structured provider contracts, whereas the studied AI2-THOR and Meta-World paths project action declarations and native observations. Both representations enter the same normalized semantic boundary.

\begin{table*}[t]
\centering\small
\setlength{\tabcolsep}{4pt}
\caption{Implemented contract and observation instantiations. Heating values are experimental configuration parameters. Timing is evaluated on the simulation's logical clock. Native simulator state supports the
benchmark evidence.}
\label{tab:contract-instances}
\par\vspace{6pt}

\begin{tabularx}{\textwidth}{@{}p{0.15\textwidth}*{3}{>{\raggedright\arraybackslash}X}@{}}
\toprule
Instance & Invocation / effect & Evidence realization & Interruption example \\
\midrule
Heating: \texttt{heat\_to}
& Target input range 20--100\,$^\circ$C; the studied target effect is 79--81\,$^\circ$C
& Three consecutive observations from a sensor provider distinct from the actuator, with identity-qualified windows, 1-logical-second spacing, maximum age of 20 logical seconds, and sample matching
& Additional observation after interrupted verification; restricted rebinding subject to its declared state constraint \\
\addlinespace
AI2-THOR: opening an object
& Retained-step ordering; native object-open condition
& Object-specific native state; projected sequencing and task-completion predicates
& Unestablished effect or stale observation must be resolved before dependent progress \\
\addlinespace
Meta-World: bounded manipulation segment
& Sequence-ready prerequisite; successor readiness or final \texttt{goal.complete}
& Native \texttt{evaluate\_state} success for final completion, with separately projected state observations
& Effect suppression and stale-observation conditions exercise verification and continuation \\
\bottomrule
\end{tabularx}

\end{table*}

Contract correctness and observation interpretation are trusted configuration. The public onboarding process checks review decisions and attestations, while the catalog rejects duplicate identities, identity drift across versions, and multiple active versions. Binding excludes non-active contracts and checks compatibility with its captured runtime context. These checks establish structural and binding consistency under the trusted semantic configuration. Appendix~\ref{app:implementation-boundaries} specifies the public-helper and runtime interfaces, version handling, and revocation behavior.


\newcounter{adfeaproposition}
\newcounter{adfeatheorem}

\newcounter{adfearule}
\newenvironment{adfearule}[2]{%
  \par\addvspace{0.45\baselineskip}%
  \Needspace{4\baselineskip}%
  \refstepcounter{adfearule}\label{#1}%
  \noindent\textbf{Rule~\theadfearule\ (#2).}\ \ignorespaces
}{\par}

\subsection{Problem Formulation: Ambiguous Outcomes and Execution Authority}
\label{sec:authority-formalization}

Let $h_t$ be a physical execution history up to time $t$, and let $x_t$ be its current state. A grounded operation $o$ identifies a task episode, an intended operation occurrence, a target, a capability, and its bound arguments. Each physical attempt has a distinct identifier $i$. A retry creates a new attempt of the same operation; renaming a tool, changing a provider, or revising a proposal does not create a new retry allowance. Intentionally repeated operations have distinct occurrences in the trusted task specification or in an admitted state-repair revision. In the latter case, a previously established state has become invalid and must be re-established for a dependent action. The repair receives a fresh occurrence and attempt identity, linked to the original grounded state-setting operation; it is not a failed attempt of the earlier successful occurrence. Let $k(o)$ be the task-episode accounting key for the grounded target, arguments, and effect semantics. The original operation, its aliases, and its state repairs share this key, so a repair does not replenish the failure-retry allowance. Parameter changes require a new authorization decision and cannot erase the unresolved outcome of an earlier attempt.

Let $\mathfrak C$ denote the available DCC capability catalog, and let $\Pi_{\mathrm{plan}}(\mathfrak C)$ denote its planner-facing view, containing capability descriptions and represented invocation/effect dependencies. The LLM uses this view, the task, and a public history $H_t$ to propose capability selections, ordering, revisions, or a completion claim. $H_t$ contains available observations, tool responses, and runtime feedback. The corresponding execution-facing contracts supply the predicates, evidence requirements, and recovery rules used by the transitions below. An effect-to-precondition dependency in a candidate plan expresses intended support; only qualified execution evidence makes that support usable progress. The physical history is not directly available to the agent or to the assurance core. A response $r_i$ reports an interaction outcome and need not determine whether an effect occurred. We distinguish the final-state predicate $\varphi(x_t)$ from an occurrence predicate $\psi(h_t,i)$. For example, observing that a door is open establishes a state predicate under suitable sensing assumptions, but does not alone establish that a particular invocation opened it. The task and contract specify which interpretation is required.

The trusted task specification contains a finite set $G$ of required obligations and their dependencies. An obligation can concern a state at completion or an event in the execution history. Proposal revision may change the candidate plan but cannot silently weaken $G$. For a DAG, verified progress is a set of supported obligations; a verified prefix is the special case obtained for a sequential serialization.

\noindent\textbf{Proposition 1 (ambiguity impossibility).}
\refstepcounter{adfeaproposition}
\label{prop:ambiguity}
Consider an operation whose duplicate physical effect is prohibited and whose provider supplies no effective deduplication or rollback. Suppose there are two admissible histories $h^+$ and $h^-$ after an attempted invocation such that: (i) the intended effect has occurred exactly once in $h^+$ and not in $h^-$; (ii) the two histories produce identical public observations until a new physical attempt is made, including any waiting or read-only observations available to the controller; (iii) a new attempt is necessary to achieve the effect in $h^-$ and would duplicate it in $h^+$; and (iv) no exogenous event resolves this distinction or establishes the missing effect. No controller restricted to that public information can guarantee all three properties: sound completion claims, absence of duplicate effects, and eventual successful completion in both histories.

\noindent\emph{Proof.} Before a new physical attempt, the controller has the same information in both histories. Soundness forbids claiming completion in $h^-$. Eventual successful completion in $h^-$ therefore requires a new physical attempt at some finite time. The same decision is made in $h^+$, where it creates a prohibited duplicate. Conversely, never retrying avoids that duplicate but violates eventual successful completion in $h^-$. Waiting or stopping does not establish the missing effect. For a randomized controller, couple its random choices in the two histories. Almost-sure successful completion in $h^-$ requires a retry with probability one and therefore violates almost-sure duplicate avoidance in $h^+$. \hfill$\square$

This is a task-specific indistinguishability argument, not a new general impossibility principle. Uncertainty about whether a failed remote call executed is a classical systems concern~\cite{birrell1984rpc}. Ambiguous outcomes and verification before retry also motivate prior tool-execution work~\cite{mansoor2026verifiedtoolcalls}. The proposition identifies when additional information or provider guarantees are necessary. ADF-EA does not resolve permanent unobservability: it may gather distinguishing evidence when available, or withhold physical recovery and successful completion otherwise.

\subsection{Evidence and Authority Transition Semantics}
\label{sec:authority-transitions}

We write the runtime state as
\[
 S_t=(G,D_t,L_t,U_t,Z_t,K_t,B_t).
\]
Here $D_t$ is the currently authorized dependency representation; $L_t$ is a retained certificate ledger; $U_t$ contains unresolved material obligations; $Z_t$ records attempt identity, settled/in-flight/unknown status, and the latest evidence verdict; $K_t$ contains the bound contract and provider context; and $B_t$ contains the remaining operation-level retry, physical-request, and observation budgets. ``Material'' means required by the task or capable of affecting its obligations or recovery policy, as declared by the trusted specification. The agent cannot remove such an outcome by dropping a node from a proposal. We distinguish $U_t^{\mathrm{exec}}$, the obligations associated with unsettled execution outcomes, from $U_t^{\mathrm{state}}$, required state predicates whose current support has been invalidated. These are tagged parts of $U_t$. A state requirement carries its dependency and required horizon: an open receptacle is needed for an outstanding Put, not forever after Put has been verified. Only requirements still material at the current decision horizon are reopened. A later closed door reopens the requirement for an open door; it does not retroactively make a previously verified Open an unsuccessful invocation. Let $\operatorname{Pending}(k,U_t^{\mathrm{exec}})$ identify unresolved execution outcomes for the accounting key $k$.

A certificate $c=(o,i,\phi,t_c,\nu,\ell)$ records an obligation predicate, its evaluation horizon, the pinned semantic revision $\nu$, and evidence lineage $\ell$. Let $\operatorname{Live}(c,t,K_t)$ mean that the certificate may still support its declared obligation at $t$. For a historical event, this requires valid historical evidence; for a current-state fact, it also requires a sound persistence argument or sufficiently current evidence. Define the usable progress view
\[
 P_t=\{c\in L_t: \operatorname{Live}(c,t,K_t)\}.
\]
Keeping an old record in $L_t$ does not keep it usable in $P_t$. Relevant expiry, context changes, or conflicting evidence must revoke its usability before a decision that relies on it. The observation and invalidation assumptions below make this temporal boundary explicit.

For a requirement associated with attempt $i$, the implemented verifier vocabulary is
\[
 q\in\{\mathrm{SATISFIED},\mathrm{UNSATISFIED},
 \mathrm{INSUFFICIENT\_EVIDENCE},\mathrm{UNKNOWN}\}.
\]
Integrity failure is handled separately and grants no authority. An effect certificate requires all applicable requirements to be satisfied and a nonempty evidence basis. An initial obligation may instead use explicitly trusted initial evidence. The runtime does not infer positive evidence from a provider acknowledgement.

Let $\operatorname{Covered}(g,P_t)$ require certificates with the correct target, operation meaning, semantic revision, and temporal interpretation. Let $\operatorname{Adm}(o,S_t)$ require valid arguments, a permitted and currently eligible binding, supported dependencies, policy compliance, and sufficient physical-request budget. Let $\operatorname{ReplayOK}(o,S_t)$ be the interruption contract's separate condition for another physical attempt. It must account for partial effects and, when relevant, a previous attempt that could still take effect. A negative final-goal predicate is not this condition.

\label{sec:selective-reexecution}
For the supported state-setting capabilities, ADF-EA separates four cases. An unresolved attempt with qualified negative evidence follows contract retry admission. A previously verified effect that has since become false follows state repair when the effect is needed by the next task action. Unknown or insufficient evidence retains the unresolved obligation and permits only allowed observation or stopping, not blind reexecution. If the relevant effect remains established, repeating that state-setting action without a distinct authorized purpose is redundant and is rejected. Execution uncertainty takes precedence: an operation with a pending outcome cannot bypass retry admission by relabeling its next request as a repair.

To formalize the second case, let $a$ be the next undispatched action in the current sequential proposal, and let $\operatorname{Link}(o,a)$ contain the current-state predicates that are both declared effects of $o$ and invocation preconditions of $a$. Links match the grounded subject, state path, value, and semantic context, not just a tool name. Write $\operatorname{Established}(o,L_t)$ for retained evidence that the relevant effect of an earlier occurrence was established. Let $\operatorname{KnownNow}(\phi,S_t)$ require a qualified, sufficiently current observation deciding $\phi$, and $\operatorname{NegNow}(\phi,S_t)$ require that this observation establishes $\neg\phi$. Define
\[
\begin{aligned}
\operatorname{RepairNeeded}(o,a,S_t) \iff{}&
 \operatorname{Established}(o,L_t)\\
&{}\land\neg\operatorname{Pending}(k(o),U_t^{\mathrm{exec}})\\
&{}\land\operatorname{Link}(o,a)\ne\varnothing\\
&{}\land\forall\phi\in\operatorname{Link}(o,a),\quad
       \operatorname{KnownNow}(\phi,S_t)\\
&{}\land\exists\phi\in\operatorname{Link}(o,a),\quad
       \operatorname{NegNow}(\phi,S_t).
\end{aligned}
\]
Current negative evidence establishes the need for repair, not failure of the historical invocation. Mere expiry of a positive certificate does not satisfy this predicate. The repair policy applies to declared state-setting effects; it does not turn an occurrence-only obligation into permission to repeat an irreversible event.

The following rules define the abstract authority boundary. Each applies when its conditions hold; the numbering provides reference identifiers, not a fixed execution sequence. A denied request is recorded without the requested physical or completion event.

\subsubsection*{Admission and evidence}

\begin{adfearule}{rule:revise-admit}{Revision and admission}
Validate a proposal against $G$, $P_t$, $U_t$, the dependency semantics, and current $K_t$. Only validated dependencies enter $D_t$. This rule creates no effect certificate, does not discard material unresolved outcomes, and does not reset operation budgets. The parent attempt and certificate history is retained when a residual proposal is admitted. Revision changes $D_t$ and its binding context, not the trusted task $G$, and creates no physical effect or evidence of one.
\end{adfearule}

\begin{adfearule}{rule:dispatch}{Dispatch}
A first attempt of $o$ requires $\operatorname{Adm}(o,S_t)$. The runtime consumes its request allowance, creates a fresh $i$, records its outstanding effects in $U_t^{\mathrm{exec}}$, and invokes the provider. Duplicate requests for the same intended occurrence use the retry rule below; a repair occurrence requires the separate state-repair rule. Neither can masquerade as an unrelated first attempt.
\end{adfearule}

\begin{adfearule}{rule:receive-qualify}{Evidence qualification}
A provider response changes interaction records and any independently justified attempt-status information. It does not itself populate $P_t$. Evidence acquisition and qualification update $Z_t$. A missing, stale, or conflicting observation does not become evidence of physical failure.
\end{adfearule}

\begin{adfearule}{rule:promote}{Progress promotion}
Qualified positive evidence for the associated obligation permits creation of a certificate in $L_t$. Resolve only material outcomes entailed by that certificate under their declared state/event meanings. Do not transfer an occurrence certificate between different attempts merely because their final-state predicates coincide. A newly verified repair can discharge the matching current-state obligation; its admission alone cannot do so. Qualified progress also discharges a dependency's temporary state requirement once its declared horizon has ended; this cannot remove a still-required goal or an unsettled execution outcome.
\end{adfearule}

\subsubsection*{Observation and recovery}

\begin{adfearule}{rule:observe}{Observation}
An allowed read-only evidence route and remaining observation budget permit reacquisition for an unresolved attempt. The rule consumes an observation allowance, makes no physical retry, and leaves the obligation unresolved until qualification supports resolution. Unknown or insufficient evidence can justify this rule or a stop; it does not by itself justify re-actuation.
\end{adfearule}

\begin{adfearule}{rule:retry-continue}{Retry and continuation}
A retry requires $\operatorname{Pending}(k(o),U_t^{\mathrm{exec}})$, a qualified negative verdict relevant to that operation, $\operatorname{ReplayOK}(o,S_t)$, a retry-permitting policy, $\operatorname{Adm}(o,S_t)$, and an unspent retry allowance. It consumes both retry and request allowances and creates a fresh attempt identifier. An ordinary successor requires $\operatorname{Adm}$ and live evidence for its dependencies. Neither route can reuse stale progress or silently reset the operation identity. Unrelated actions may proceed if their own dependencies and the policy permit them; the implementation studied below uses sequential proposals.
\end{adfearule}

\begin{adfearule}{rule:state-repair}{State repair}
Require $\operatorname{RepairNeeded}(o,a,S_t)$, a policy permitting re-establishment of the declared state-setting effect, and $\operatorname{Adm}(o,S_t)$. The policy must cover the repair's other effects and exclude an outstanding attempt that could still take effect. Admit a residual proposal beginning with the repair and continuing toward the original task obligations. Every actual dispatch still undergoes current invocation, binding, and budget checks; admission of the residual plan does not authorize its suffix in advance. In particular, unresolved prerequisites cannot be skipped by shortening the proposal.

\Needspace{9\baselineskip}
At revision admission, preserve $L_t$, $Z_t$, $U_t$, and all unspent budgets. Dispatch creates a fresh repair attempt, adds its outstanding effects to $U^{\mathrm{exec}}$, and consumes one physical-request unit. If $b_p$, $b_v$, and $b_r(k)$ denote remaining physical, observation, and retry allowances immediately before dispatch, then
\[
 (b'_p,b'_v,b'_r(k(o)))=(b_p-1,b_v,b_r(k(o))).
\]
Thus repair uses physical budget but no failure-retry allowance, and does not reset the latter. An unsuccessful or uncertain repair creates a new unresolved outcome; subsequent requests must use the ordinary retry/observation rules. Only qualified positive evidence can re-establish usable progress and resolve the corresponding obligations. Still-live independent certificates remain available; facts affected by the repair require revalidation.
\end{adfearule}

\subsubsection*{Rejection and termination}

\begin{adfearule}{rule:reject-repetition}{Redundant or unsupported repetition}
Reject a repeated state-setting request when it has no pending execution outcome and all linked effects remain established, or when no required downstream dependency justifies its repetition. Reject repair admission when the required current evidence is unknown or insufficient. This rule concerns the represented state-setting task, not a distinct intentionally repeated event authorized by another contract.
\end{adfearule}

\begin{adfearule}{rule:terminal}{Invalidation, stopping, and completion}
Invalidated certificates remain historical records but are excluded from $P_t$; material current-state obligations they no longer support enter $U_t^{\mathrm{state}}$. Invalidation preserves the historical success record and does not by itself create a failed-attempt obligation in $U_t^{\mathrm{exec}}$. Stopping grants no completion authority. Authorize a completion claim only if
\[
 \left(\forall g\in G,\ \operatorname{Covered}(g,P_t)\right)
 \land U_t=\varnothing
 \land \operatorname{TerminalPolicyOK}(S_t).
\]
The terminal policy also excludes unresolved in-flight effects that could invalidate the claimed task result.
\end{adfearule}

\subsection{Conditional Soundness}
\label{sec:conditional-soundness}

We use the following assumptions. (A1) The trusted task and effect predicates correctly describe the stated obligations, including target, arguments, event versus state meaning, and semantic version. This includes the state-setting repair policy, grounded effect--precondition links, and the accounting key shared by retries, aliases, and repairs. (A2) Positive evidence accepted by qualification is sound for the associated predicate at its evaluation horizon; negative evidence used for recovery is sound for the relevant negative predicate. This includes the required source and execution correlation assumptions. For state repair, the accepted current observation is sound for the linked state predicate and does not substitute a historical success or an unknown value for a current negative observation. (A3) The initial certificates are sound, and $\operatorname{Live}$ is sound at every authorization point: relevant invalidation is processed before use, or a provider-level atomic check/persistence guarantee covers the observation-to-action interval. (A4) All relevant dispatches and completion events are mediated by the rules above; state updates and budget consumption are serialized, and trusted state cannot be overwritten by agent output. Residual revision retains the original material obligations, attempt history, and unspent budgets; only its admitted child remains active for further dispatch. (A5) The declared coverage relation is sound: covering every required obligation implies the specified task goal. These assumptions concern the actual deployment, not merely schema validity.

\noindent\textbf{Theorem 1 (completion soundness).}
\refstepcounter{adfeatheorem}
\label{thm:completion-soundness}
Under A1--A5, every completion authorized by the transition system at time $t$ satisfies the trusted task obligations in $h_t$ with their declared temporal interpretations. In particular, an obligation requiring a current-state predicate holds in $x_t$, not merely at a past horizon.

\noindent\emph{Proof.} Use induction over transitions to maintain the invariant that each certificate in $P_t$ soundly supports its associated obligation at $t$. The initial case follows from A3. Revision, dispatch, responses, and stop do not add evidence-supported progress. In particular, state-repair admission does not restore a revoked certificate: its current-state obligation remains unresolved, and repair dispatch adds an execution obligation. Promotion adds a sound certificate by A1--A2. Observation alone adds no certificate. Invalidation and the definition of $P_t$ remove unusable certificates; retained certificates remain sound by A3, including across physical actions. Repair success can restore current-state support only through this same promotion rule; it cannot reuse the earlier occurrence's certificate as evidence of the new attempt. By A4, a residual child retains the parent's material obligations and cannot authorize completion merely because its shorter suffix has ended. A4 excludes other state-changing authorization paths. At completion, every required obligation is covered by this invariant, and A5 gives the stated task goal. The unresolved-outcome and terminal-policy checks prevent completion while material execution uncertainty remains. \hfill$\square$

\noindent\textbf{Theorem 2 (recovery authorization soundness).}
\refstepcounter{adfeatheorem}
\label{thm:recovery-soundness}
Under A1--A4, every authorized physical reexecution in the represented state-setting task satisfies exactly one of two admission cases:
\begin{enumerate}
\item A \emph{contract retry} has a correlated unresolved operation, qualified negative evidence, explicit replay permission, admissible current binding and dependencies, and unspent physical and retry allowances.
\item A \emph{state repair} has retained establishment evidence, no pending execution outcome for its accounting key, a qualified current negative observation of an effect required by the next action, an admitted residual proposal, and a policy-permitted binding with unspent physical allowance.
\end{enumerate}
Unknown or insufficient evidence alone authorizes neither case; an already-established linked state does not justify redundant repair. Revision, provider aliasing, and repair admission cannot replenish any budget or discard an unresolved material outcome. Repair does not consume failure-retry allowance, but every repair dispatch consumes physical budget. Observation-only recovery issues no physical reexecution, and still-live certificates remain available across proposal revision.

\noindent\emph{Proof.} Induct over transitions, maintaining the operation-to-attempt mapping, monotonically non-increasing allowances, retained history, and preservation of unresolved material obligations until evidence-supported resolution. Revision preserves these invariants. Retry and state repair are the only rules that admit physical reexecution: the former requires $\operatorname{Pending}(k,U^{\mathrm{exec}})$ and the latter its negation, so their admission cases are disjoint. The remaining premises of each rule establish the corresponding evidence, policy, and dependency conditions. Retry decreases both physical and retry allowances; repair decreases only the physical allowance, preserving the retry count and shared key. A4 serializes these changes and prevents both parent and child from spending the same residual budget. A repair's fresh attempt identity therefore grants no fresh allowance. Both rules introduce outstanding execution obligations; neither clears earlier ones by renaming the plan. Unknown evidence satisfies neither the qualified-negative retry premise nor the current-negative repair premise. Observation and rejection produce no physical event, and revision retains the ledger entries supporting still-live progress. \hfill$\square$

This theorem establishes policy-relative authorization, not unconditional physical safety or exactly-once execution. To infer absence of duplicated irreversible effects, an additional replay-soundness assumption is needed: every replay admitted by $\operatorname{ReplayOK}$, and every additional effect of an admitted state repair, must be physically safe in all histories consistent with its evidence, including delayed completion and partial-effect histories. Preservation of current prefix facts across recovery also requires the recovery action's frame conditions or fresh evidence. A policy permitting retry does not prove either assumption.

The evaluated sequential adapter realizes the four-way decision through \texttt{dispatch}, \texttt{retry\_available}, and \texttt{\_state\_repair\_required}. Contract retry uses Boolean qualified negative readback and the existing recovery ledger. State repair instead requires fresh public facts, a declared effect linked to the next undispatched action's precondition, a currently false linked predicate, and an authorized invocation. A fresh residual lifecycle retains the parent artifacts, shares the same recovery ledger, inherits the remaining physical budget, and uses the common observation budget. New repair effects pass through ordinary verification. Completion first checks unresolved ledger obligations and then the active lifecycle's progress and final evidence.

This mapping is narrower than automatic repair of arbitrary DAG dependencies. The repair helper matches projected state paths and values in a trusted task context and relies on the adapter's freshness and subject grounding; it is not a separate sensor-authentication or temporal-validity verifier. $U^{\mathrm{state}}$ is represented by current public precondition facts and progress checks, rather than a second persistent ledger collection. Repair identity is recorded through the revision reason and fresh execution correlation, while retry accounting remains keyed to the grounded effect. The theorems specify the abstract rules and their assumptions, not a proof that every Python path refines them. The supported synchronous state-setting experiments establish the reported behavior; asynchronous partial effects, arbitrary dependency invalidation, and concurrent budget use require their own implementation guarantees.

\subsection{Realizing the Agent--Execution Loop}
\label{sec:runtime-lifecycle}

The runtime resolves capability identifiers, arguments, targets, and proposed ordering against the DCC catalog and task specification. It materializes the represented dependencies into an inspectable execution DAG and checks grounding, invocation constraints, budgets, and provider eligibility. The binding identifies the concrete tool and evidence routes. A residual proposal undergoes the same checks while retaining the original obligations and remaining budget.

Provider responses retain invocation lineage; effect observations supply the verification basis. Qualification checks the bound requirement, source, execution and target association, and applicable freshness and observation conditions. It yields satisfied, unsatisfied, insufficient-evidence, or unknown requirements. Promotion requires a nonempty qualifying basis satisfying all effect requirements. A failed identity or content check does not by itself establish physical failure: retry requires evidence of the relevant negative effect, not merely an unsuccessful verification procedure.

The active residual lifecycle shares the recovery ledger and inherits the remaining physical budget. Completion checks unresolved obligations and required task evidence. Public feedback exposes progress and authorization decisions to the agent, which chooses its next action or explicitly stops. Retry and state repair follow the implementation mapping in Section~\ref{sec:conditional-soundness}.

Appendix~\ref{app:execution-lifecycle-details} preserves the detailed qualification procedure, implementation interfaces, and the Heating continuation example. The evaluated sequential state-repair mapping is specified with the formal model above.

\section{Evaluation}
\label{sec:evaluation}

The evaluation tests execution assurance given explicit task goals, constraints, and capability contracts. \textbf{RQ1:} Can the contracts support composition execution checks while preserving permitted completion? \textbf{RQ2:} Do obligations, evidence requirements, and permissions persist through faults, recovery, and proposal revision? \textbf{RQ3:} Can fixed contracts support new compositions? The primary controlled comparison addresses RQ1--RQ2; a frozen-contract study addresses RQ3. Native execution cases, internal ADF ablations, and domain/model/framework integrations provide complementary evidence without pooling their populations.

The primary comparison uses a local Boolean Device simulator with explicit dependency predicates, correlated evidence, and recorded native events. The supporting environments exercise complementary requirements: Heating combines actuation, measurement, and recording; Room-exit links light, door, and lock actions; AI2-THOR provides household object-state dependencies; Meta-World exposes manipulation skills backed by continuous control. All environments are simulated. Domain contracts and observation adapters bind their semantics to the execution lifecycle.

The external baseline comparison uses the public execution mechanisms of ToolGate Algorithm~1 and Verified Tool Calls (VTC) Algorithm~1; Direct supplies a forwarding reference. The comparison comprises 68 executions with frozen method implementations and independent event-based scoring. The baselines retain their published state, checks, verification, and retry behavior. Their domain adapter binds simulator I/O and the required predicate inputs without adding recovery, cross-request obligations, budget enforcement, or completion gates. Section~\ref{sec:latest-controlled-comparison} states the evaluated algorithm versions and input bindings. The model/framework studies compare ADF-EA with policy-informed Direct; ADF component removals are reported separately as internal ablations.

We report goal-confirmed completion and evidence-qualified completion separately from dispatch violations and the supplementary measure of violation-free completion. A necessary state repair restores an invalidated prerequisite; improper repetition performs additional physical work without the corresponding authority. Agent requests, provider dispatches, observations, state reads, and native controller steps are counted separately. Completion denominators include only conditions designated for completion; false-completion and violating-run denominators include all scheduled conditions. Table~\ref{tab:latest-controlled-outcomes} presents the primary results; Appendix~\ref{app:latest-controlled-details} gives recovery, call counts, and independent-scoring details. Supporting studies use the outcome definitions and populations specified in their respective sections.

\subsection{Controlled Comparison with Published Algorithms}
\label{sec:latest-controlled-comparison}

The experiment contains 17 controlled conditions on three task identifiers, covering a dependency chain and conjunctive completion requirements. Four methods run once per condition, yielding 68 formal executions; a separate four-condition pilot yields 16 executions and is excluded from the main results. All runs are local and make zero model calls. The 17 conditions are fault/request configurations, not 17 independent tasks. Two conditions use normal legal requests. The remaining conditions cover missing, stale, wrong-object, wrong-source, and wrong-correlation evidence; an invalidated prerequisite followed by missing evidence; an unknown outcome across proposal revision; retry and shared-budget probes; omission of an unresolved obligation; provider replacement; redundant requests; and ambiguous responses with or without a physical effect. Fourteen conditions are designated for completion, while three are stopping probes. These groups are fixed in the shared configuration and contain the same condition IDs for all methods. A stopping probe tests the response to its supplied requests; the label does not establish that no legal alternative continuation exists.

Task goals and budgets, capability preconditions/effects and evidence rules, and candidate action arrangements are represented separately. Each method starts from its own identical initial simulator state and receives the same original candidate-request stream and external fault rules. Observations and faults are evaluated against that method's actual actions and event history; a blocked action does not advance another method's state. Method-internal verification, observation, and retry remain enabled and are counted. Consequently, the common original requests do not imply identical executed traces, observation counts, or submitted suffixes after termination.

ToolGate implements the precondition, invocation, and postcondition/commit logic from Section~3.2 and Appendix~E Algorithm~1 of its public paper~\cite{liu2026toolgate}. The controlled request identifies a single candidate, so retrieval is a singleton binding. VTC implements the printed control flow of Algorithm~1 with retry bound $N=1$ and reuses its idempotency key~\cite{mansoor2026verifiedtoolcalls}; the shared simulator provides no backend idempotency service. These are our reproductions of the published execution algorithms, not executions of the authors' complete software stacks. A common Boolean predicate binding supplies the domain inputs. No additional cross-request obligation gate, task-completion gate, or recovery policy is added to either algorithm. ADF-EA uses its existing execution loop with a contract-authorized retry hook: the hook can request another attempt only when the runtime reports retry authority, and it retains the original request stream and shared budgets.

\begin{table*}[!htbp]
\centering\small
\setstretch{1}
\setlength{\tabcolsep}{5pt}
\renewcommand{\arraystretch}{1.12}
\caption{Primary controlled comparison: 17 conditions per method, with 14 completion-permitted conditions and three stopping probes. ToolGate and VTC denote frozen public-algorithm reproductions. Evidence-qualified completion evaluates the accepted endpoint; earlier dispatch violations remain separately counted. Incorrect blocking uses each method's actually received legal physical requests as its denominator.}
\label{tab:latest-controlled-outcomes}
\begin{tabularx}{\textwidth}{@{}>{\raggedright\arraybackslash}X rrrr@{}}
\toprule
Metric & ADF-EA & ToolGate & VTC & Direct \\
\midrule
Goal-confirmed completion (permitted conditions) & 14/14 & 9/14 & 14/14 & 13/14 \\
Evidence-qualified completion (permitted conditions) & 14/14 & 9/14 & 12/14 & 11/14 \\
Accepted completion with physical goal unmet (runs) & 0/17 & 0/17 & 1/17 & 2/17 \\
Accepted completion: goal met, evidence unqualified (runs) & 0/17 & 0/17 & 3/17 & 3/17 \\
Violating provider dispatches & 0 & 10 & 14 & 14 \\
Runs containing a violating dispatch & 0/17 & 8/17 & 11/17 & 11/17 \\
Unavailable-capability dispatches & 0 & 1 & 1 & 1 \\
Improper repeated dispatches & 0 & 7 & 11 & 10 \\
Legal physical requests incorrectly blocked & 0/53 & 0/42 & 0/49 & 0/49 \\
Safe terminal stop (stopping probes) & 3/3 & 2/3 & 0/3 & 0/3 \\
Unnecessary stop (permitted conditions) & 0/14 & 5/14 & 0/14 & 0/14 \\
\bottomrule
\end{tabularx}
\end{table*}

ADF-EA completes all 14 permitted conditions with qualified evidence and incurs no violating provider dispatches. ToolGate, VTC, and Direct complete 9, 12, and 11 with qualified evidence, respectively. VTC also reaches 14/14 goal-confirmed completion, showing why evidence qualification and dispatch legality must be evaluated separately from physical goal attainment. The three comparison methods incur 10, 14, and 14 violating dispatches and 7, 11, and 10 improper repetitions. All four methods record zero incorrect blocks of legal physical requests. ToolGate terminates unnecessarily in five completion-permitted conditions, so its lower provider-call count accompanies reduced completion.

The supplementary requirement of a qualified endpoint with no earlier violating dispatch yields 14/14, 3/14, 4/14, and 4/14 for ADF-EA, ToolGate, VTC, and Direct. These are \emph{violation-free completion} counts, not the evidence-qualified completion counts in the main table. For example, all four methods eventually obtain qualified completion in the stale-evidence condition, although the comparison methods also execute an improper repeat before reacquiring evidence.

Physical goal attainment is reported separately for completion-designated conditions and stopping probes. Over the fourteen completion-designated conditions, final physical goal attainment is 14/14 for ADF-EA, 10/14 for ToolGate, 14/14 for VTC, and 13/14 for Direct, without requiring a completion declaration. The all-condition physical totals also include three stopping probes: VTC's additional physical attainment relative to ADF-EA comes from executing beyond the shared task budget in the budget-reset probe. Table~\ref{tab:latest-controlled-extra} reports each group separately and the aggregate final-state outcome.

Recovery is useful as well as restrictive: ADF-EA obtains qualified completion in all seven evidence-recovery conditions, all three retry-permitted conditions, and the single condition combining state repair with later evidence loss. It dispatches 53 physical calls, compared with 52, 65, and 63. Appendix~\ref{app:latest-controlled-details} reports the call and request counts.

\subsection{Composition Reuse after Freezing Contracts}
\label{sec:frozen-composition-reuse}

A separate study freezes seven capabilities, their contracts, the Boolean domain mapping, and the generic runtime before defining six new compositions. The task definitions change the use of initial-state dependencies, interleave branches, preserve a completed branch, or combine a positive goal with preservation of a false-valued fact. They reuse existing capabilities without modifying contracts, the mapping, or execution code for individual cases. Candidate arrangements remain explicit evaluation inputs; this experiment tests execution support for new compositions, not autonomous plan discovery.

ADF-EA supports five of the six compositions. The sixth, F4, requires completing one branch while preserving a false fact on another. Although the preservation fact already holds initially, the frozen ADF mapping incorrectly requires a producer for it and fails during initialization. This is a mapping failure and remains in the denominator. ToolGate, VTC, and Direct complete all six in these fault-free traces. The six previously used composition structures pass 6/6 for every method but are development regressions, not additional held-out successes. This study uses a separate frozen runtime version, specified in Appendix~\ref{app:latest-controlled-details}, and its runs are excluded from the 68-run controlled comparison. Appendix~\ref{app:latest-controlled-details} lists the new structures and the unchanged implementation boundary.

\subsection{Native Execution Cases: State Repair and Evidence Loss}
\label{sec:native-agent-mechanism-comparison}

An AI2-THOR development pilot evaluates six ADF-EA episodes: three task instances under two conditions, with one episode per combination. The instances require ToiletPaper inside Cabinet in FloorPlan414, CreditCard inside Box in FloorPlan304, or SoapBottle inside Cabinet in FloorPlan27, with the receptacle closed at completion. These are execution case studies using the existing ADF runtime, separate from the public-baseline experiment in Section~\ref{sec:latest-controlled-comparison}.

The ReAct agent selects pickup, open, put, close, and read-only observations from public feedback. Each episode allows eight physical requests, four observations, and twelve model calls. The requested model is \texttt{deepseek-v4-flash}, with thinking disabled, temperature zero, and a 1,024-token output limit; the returned identifier is \texttt{deepseek-flash}, recorded without assuming a verified alias. Fault labels and evaluator state are not supplied to the agent. Native goal satisfaction, evidence qualification, and completion authorization are recorded separately.

\begin{table*}[!htbp]
\centering\small
\setstretch{1}
\setlength{\tabcolsep}{5pt}
\renewcommand{\arraystretch}{1.12}
\caption{ADF-EA native execution cases: three episodes per row, six in total. These are development observations. Necessary state repairs and rejected requests are excluded from improper repeated dispatches.}
\label{tab:native-agent-mechanism-comparison}
\begin{tabularx}{\textwidth}{@{}>{\raggedright\arraybackslash}p{0.21\textwidth} >{\raggedright\arraybackslash}X rrrr@{}}
\toprule
Condition & Terminal outcome & Dispatches & Repairs & \shortstack{Improper\\repeats} & \shortstack{Model\\calls} \\
\midrule
Container state invalidated & 3/3 qualified, authorized completions & 15 & 3 & 0 & 27 \\
Put evidence unavailable & 3/3 explicit agent stops; goal unmet & 9 & 0 & 0 & 27 \\
\bottomrule
\end{tabularx}
\end{table*}

In the repair condition, an external intervention closes a previously opened and verified receptacle before placement. ADF-EA rejects premature Put requests and admits a necessary Open through a state-repair residual proposal, preserving the ledger and remaining budget. All three tasks recover with one repair each and complete with qualified evidence. The historical Open outcome remains settled, while a current observation falsifies the prerequisite for Put; Rule~\ref{rule:state-repair} admits restoring that prerequisite.

In the evidence-loss condition, Put physically succeeds but its effect evidence remains unavailable. ADF-EA retains the unresolved obligation and intercepts all four blind reexecution requests across the three trajectories. Each trajectory dispatches only pickup, open, and put, with no improper repeated execution. The agents explicitly stop after eight, eight, and eleven model calls. Close remains withheld, so the overall task goal is unmet. None submits an unsupported completion claim. These ADF-EA case studies demonstrate legal state repair and retention of unresolved execution obligations.

\subsection{Mechanism Evidence: Observation, Retry, and Continuation}
\label{sec:mechanism-evidence}

Internal ADF ablations test why the runtime can continue useful work without repeating actions whose outcomes remain uncertain. Removing evidence reacquisition loses all fourteen stale-evidence recoveries in the original faulted population; removing continuation authorization restores all fourteen unavailable-successor dispatches. These experiments remove ADF components and preserve the full system as the reference (Section~\ref{sec:eval-attribution}).

Three fixed-request ADF-EA runs in FloorPlan425 test normal execution, temporary placement-evidence loss, and persistent placement-evidence loss. The task requires placement inside the receptacle and a closed receptacle at completion. Each run allows eight physical requests and four read-only observations, with no model calls. Put executes in both fault conditions, but its action-bound evidence is withheld. Two scripted observations follow: the temporary condition reveals qualified evidence on the second observation, whereas the persistent condition keeps both unknown. The release rule depends on condition and observation ordinal; native state is reserved for scoring.

ADF-EA completes the normal run with one Put dispatch and no extra observations. Under temporary evidence loss, two observations discharge the Put obligation in the DCC recovery ledger, enabling Close and qualified completion without replay. Under persistent evidence loss, the later repeated Put request is rejected: two Put requests produce only one dispatch. Completion is also rejected because the placement evidence remains unresolved and the receptacle remains open.

Paired recovery-policy controls hold the missing-effect fault fixed while changing retry permission. With a retry allowance, ADF-EA establishes the effect and authorizes completion; without permission, it rejects the repeated request. The 44-run control study preserves all completion-permitted outcomes and separates legal physical recovery from observation-only recovery (Section~\ref{sec:eval-policy-controls}). Together with the state-repair cases above, these cases exercise the distinction between an unresolved failed invocation and a previously verified state that later becomes invalid.

The lifecycle and semantic-contrast studies test the breadth of these decisions. Across 49 cases, ADF-EA selects every prescribed response---completion, observation, rejection, or stopping---versus eight for Direct. Twelve paired contrasts then vary invocation, effect, evidence, or interruption semantics. Intact contracts and sham mutations retain all twelve required distinctions, while targeted weakening retains none across the 72 executions. Each DCC block therefore contributes to a distinct authorization or verification decision. Appendix~\ref{app:mechanism-controls} gives the contrasts and ablations.

\subsection{Cross-Domain Skill Reuse}
\label{sec:cross-domain-skill-reuse}
\label{sec:metaworld-mechanism-comparison}

Two ADF-EA runs in native Meta-World \texttt{pick-place-v3} examine the contract boundary around robotic skills. The task uses seed~0 and the official \texttt{SawyerPickPlaceV3Policy}, exposed as approach, grasp-and-lift, and transport skills. Each run allows eight high-level requests, four additional observations, and 500 controller steps. Normal execution and persistent evidence insufficiency use the same controller; no model is called.

\begin{table*}[!htbp]
\centering\small
\setstretch{1}
\setlength{\tabcolsep}{5pt}
\renewcommand{\arraystretch}{1.12}
\caption{ADF-EA skill execution cases. Provider calls count high-level skill dispatches; controller steps count native MuJoCo execution. One run per condition.}
\label{tab:metaworld-mechanism-comparison}
\begin{tabularx}{\textwidth}{@{}>{\raggedright\arraybackslash}X c c r r r@{}}
\toprule
Condition & \shortstack{Native\\goal} & \shortstack{Qualified\\completion} & \shortstack{Provider\\calls} & \shortstack{Repeated\\calls} & \shortstack{Control\\steps} \\
\midrule
Normal & Met & Yes & 3 & 0 & 52 \\
Persistent evidence insufficiency & Met & No & 3 & 0 & 52 \\
\bottomrule
\end{tabularx}
\end{table*}

Normal execution completes with qualified evidence. In the evidence-insufficient run, the final skill executes, but its effect evidence and one subsequent read-only observation remain unavailable. ADF-EA rejects a repeated transport request before entering the provider interface and withholds completion. The simulator-native physical goal is satisfied in both runs, with three skill calls and 52 controller steps, while evidence-qualified completion differs. This demonstrates the represented execution rule at a skill interface; the controller remains responsible for the underlying continuous behavior.

\subsection{Execution Assurance across Models and Agent Frameworks}
\label{sec:adaptive-system-results}

The model and framework studies test execution assurance through both autonomous interaction and execution of model-generated proposals. Adaptive Room-exit and AI2-THOR ReAct agents choose actions and stopping from feedback; a separate four-model study supplies proposals for paired replay; five installed agent frameworks exercise their native model/tool loops. These designs vary the upstream model and integration interface while retaining the same execution rules. Table~\ref{tab:adaptive-system-summary} summarizes the adaptive comparisons. Within each adaptive study, both arms receive the declared capabilities, policy, and public observations. ADF-EA additionally enforces authorization and maintains execution state.

\begin{table*}[t]
\centering\small
\setlength{\tabcolsep}{5pt}
\renewcommand{\arraystretch}{1.12}
\caption{Adaptive execution assurance with preserved completion. Completion denominators cover conditions that permit completion. The other rows report false-completion fractions or counts of unavailable-provider dispatches. GLM outcome metrics use 18 valid Direct and 16 valid ADF-EA trajectories; two interrupted ADF-EA runs are retained separately in Table~\ref{tab:eval-room-exit}.}
\label{tab:adaptive-system-summary}
\begin{tabularx}{\textwidth}{@{}>{\raggedright\arraybackslash}p{0.27\textwidth}>{\raggedright\arraybackslash}X rr@{}}
\toprule
\textbf{Adaptive study} & \textbf{Outcome} & \textbf{Direct} & \textbf{ADF-EA} \\
\midrule
\textbf{Room-exit: Flash} & Feasible-task completion & 9/9 & \textbf{9/9} \\
& False completion & 6/18 & \textbf{0/18} \\
& Unavailable dispatches & 7 & \textbf{0} \\
\addlinespace[4pt]
\textbf{Room-exit: GLM MAX} & Feasible-task completion & 9/9 & \textbf{9/9} \\
& False completion & 6/18 & \textbf{0/16} \\
& Unavailable dispatches & 8 & \textbf{0} \\
\addlinespace[4pt]
\textbf{AI2-THOR ReAct} & Permitted qualified completions & 25/25 & \textbf{25/25} \\
& Unavailable dispatches & 10 & \textbf{0} \\
\addlinespace[4pt]
\textbf{Five agent frameworks} & Permitted qualified completions & 50/50 & \textbf{50/50} \\
& Unavailable dispatches & 50 & \textbf{0} \\
\bottomrule
\end{tabularx}
\end{table*}

Room-exit tests the distinction between a correct endpoint and an authorized execution trace. Both DeepSeek V4.1 Flash and GLM-5.3 MAX complete all nine feasible runs in each arm. With each model, Direct also declares completion in six runs whose physical goal is unmet; ADF-EA records none in valid trajectories. It blocks the seven unavailable calls made by Flash Direct and the eight made by GLM Direct. Repeated dispatches fall from twelve to zero with Flash and from five to zero with GLM. Flash safely terminates all nine infeasible runs; GLM explicitly stops in all seven valid infeasible runs. The two GLM transport interruptions and all terminal sources appear in Tables~\ref{tab:eval-room-exit} and~\ref{app:tab:room-exit-terminals}.

The AI2-THOR ReAct reevaluation tests recovery utility through 15 normal and 15 faulted tasks per arm. ADF-EA preserves all 25 permitted completions, including five authorized retries and five evidence reacquisitions, while preventing all ten unavailable-provider calls made by Direct. It incorrectly blocks none of its 86 legal requests. The five-framework study extends this behavior through OpenAI Agents SDK, LangGraph, Google ADK, Microsoft Agent Framework, and smolagents, each running its installed model/tool loop. Across 150 runs, ADF-EA preserves all fifty permitted completions, prevents fifty unavailable calls, and incorrectly blocks none of 197 legal requests. Both arms stop in every no-continuation run; ADF-EA additionally enforces the policy throughout the preceding dispatches. All runs in these two studies finish without protocol, transport, or budget failure (Tables~\ref{tab:eval-adaptive} and~\ref{tab:eval-frameworks}).

A separate proposal-and-paired-replay study tests the same execution rules with proposals from Qwen3.8 Max, DeepSeek V4 Pro, Kimi K3, and DeepSeek V4.1 Flash. All sixty proposals are valid. For each model, ADF-EA preserves five goal-confirmed completions and reduces both false completions and runs with violating actions from five to zero. The adaptive and replay results together show that the assurance rules remain effective across different upstream models and integration interfaces (Appendix~\ref{app:cross-model}).

The interaction results identify the cost of these decisions. Counts below are Direct versus ADF-EA. ReAct uses 129 versus 122 model calls, and the five-framework study uses 351 versus 297, with additional metadata reads supporting runtime checks. Room-exit instead requires more model interaction to contain repeated proposals: Flash uses 89 versus 118 calls; GLM uses 81 versus 113 in valid trajectories, or 131 ADF-EA calls including its interrupted runs. Table~\ref{tab:eval-adaptive-cost} and Section~\ref{sec:eval-cost-boundaries} report the separate call, observation, and dispatch counts. These studies demonstrate preserved task utility and stronger execution control across the supported integrations.

\section{Discussion}
\label{sec:discussion}

ADF-EA connects agent autonomy to persistent execution authority. The agent can revise its proposal as conditions change, while verified progress, unresolved outcomes, and remaining budgets continue to govern the task. DCC supplies the common capability semantics through which planning, evidence interpretation, and execution authorization interact. The result is an architecture that supports adaptation while making the consequences of earlier actions operationally binding on later decisions.

The experiments demonstrate complementary benefits of this architecture. State repair restores invalidated prerequisites and preserves completion. Observation recovery establishes missing evidence without repeating physical work. Persistent obligations block unsupported reinvocation, and current admission checks prevent unavailable calls even when an agent repeatedly proposes them. The adaptive model and framework results show that these controls coexist with full retention of completion in the permitted conditions. Execution assurance therefore supports useful autonomy through both recovery and restraint, selected by the task's actual execution state.

The formal and empirical results play complementary roles. The soundness theorems establish properties of the declared transition rules, including obligation retention and distinct retry/repair admission. The experiments show how the implemented architecture uses those distinctions during execution: the agent can restore a lost prerequisite, but cannot obtain another dispatch merely by repeating a request whose outcome remains unresolved. The research contribution is the reusable execution mechanism and its demonstrated behavior; the proofs make its authorization properties explicit.

The reuse results show how common semantics accommodate different execution interfaces. Meta-World places admission above a skill containing multiple controller steps; AI2-THOR places it above discrete household actions; Heating connects actuation to measurement and recording. The four-model replay study and five SDK integrations vary the upstream proposal and interaction layers. Domain contracts and adapters provide the predicates, evidence routes, and recovery permissions, while the core retains the same rules for progress, obligations, renewed authority, and completion. This is the architectural value of DCC: capability knowledge becomes reusable execution semantics across the planning--execution boundary.

\subsection{Deployment Scope and Further Directions}
\label{sec:limitations}

The formal properties depend on correct contract semantics, sound and timely evidence, and mediated execution. The implemented state-repair path handles represented sequential state-setting tasks through trusted public-state projections. Asynchronous partial effects, concurrent budget use, and arbitrary dependency invalidation require corresponding provider and runtime guarantees. Real-device deployment additionally requires reliable sensing and protection across observation-to-action intervals; continuous control remains the skill Provider's responsibility.

The experiments specify their task populations, contract policies, model configurations, and integration paths. Extending the architecture to additional task families and real-device deployments will further test its sensing, timing, and provider guarantees.

\section{Conclusion}
\label{sec:conclusion}

ADF-EA unifies capability-based planning and evidence-grounded execution through Device Capability Contracts and persistent execution state. Its lifecycle supports observation recovery, permitted retry, and necessary state repair while retaining the obligations and budgets that govern continued action. Conditional soundness results characterize completion and recovery authorization. Evaluations across multiple LLMs, five agent frameworks, and simulated device domains demonstrate complementary benefits in execution assurance. The controlled comparison with public-algorithm reproductions of ToolGate and VTC and with Direct yields evidence-qualified completion in all 14 completion-designated conditions and zero violating dispatches for ADF-EA. Autonomous household case studies complete all three required state repairs and record no improper repetitions under persistent evidence loss. The framework integrations preserve all fifty permitted completions while preventing all fifty unavailable-provider calls made by Direct. Controlled comparisons and ablations explain how the shared execution rules produce these gains. Together, the results establish a reusable architecture for agents to plan, act, and recover under consistent execution rules.

\printbibliography
\clearpage
\appendix

\section{Supporting Evaluation Studies}
\label{app:supporting-evaluations}

This appendix explains the supporting studies behind the main results. Each study identifies the execution question, the compared methods, and the behavior established by its outcomes. Adaptive studies test model-driven execution and recovery; fixed-proposal controls and contract perturbations identify the mechanisms; composed faults and domain cases test retained progress; framework and model integrations test reuse. Condition-level tables preserve the original study populations and event counts.

\subsection{Controlled Comparison: Scoring and Reproducibility}
\label{app:latest-controlled-details}

The controlled comparison comprises 68 formal executions and a separate 16-execution pilot, with no model calls. Its artifacts are archived as \texttt{paper\_metrics\_20260921}. The input, freeze manifest, original candidate streams, fault schedules, all raw events, independent scores, and summaries are retained. Three task identifiers are present: \texttt{A\_chain}, \texttt{A\_conjunction}, and \texttt{dev\_chain\_conjunction}; the latter two share conjunctive goal structure. Normal requests, fault conditions, and response/continuation probes are not independent task samples. The retry and evidence groups below can overlap, so their denominators must not be added.

The scorer reconstructs physical state at each accepted completion claim from native simulator events and separately evaluates the task goal. It checks evidence subject, source, type, timestamp, execution correlation, and value against the independently retained event history and experimental contract requirements. It does not use the runtime's admission or qualification verdict as its answer. It also reconstructs request-time legality, available providers, relevant operation history, and consumed budgets. Final physical goal attainment without a completion claim, evidence-qualified completion after an earlier violation, and a safe stopping endpoint after an earlier violation remain distinct outcomes. Seven scorer tests cover evidence-field mutations, missing evidence, unclaimed physical success, prior violations, and native-state/event inconsistency; nine evidence-field mutations are included. These synthetic scorer checks are not experimental task samples.

\begin{table*}[!htbp]
\centering\small
\setstretch{1}
\setlength{\tabcolsep}{5pt}
\renewcommand{\arraystretch}{1.12}
\caption{Recovery outcomes for the 17-condition controlled comparison. Groups overlap. The three no-external-fault conditions include a duplicate-request probe; they are not all legal request traces. A necessary repair dispatch is counted separately from successful recovery of the entire task.}
\label{tab:latest-controlled-recovery}
\begin{tabularx}{\textwidth}{@{}>{\raggedright\arraybackslash}X rrrr@{}}
\toprule
Metric & ADF-EA & ToolGate & VTC & Direct \\
\midrule
No external fault: qualified completion & 3/3 & 3/3 & 3/3 & 3/3 \\
Evidence-recovery conditions: qualified completion & 7/7 & 4/7 & 5/7 & 5/7 \\
Stale-evidence subset: qualified completion & 1/1 & 1/1 & 1/1 & 1/1 \\
Retry-permitted conditions: qualified completion & 3/3 & 0/3 & 2/3 & 1/3 \\
State-repair condition: qualified completion & 1/1 & 0/1 & 0/1 & 0/1 \\
Necessary state-repair dispatches & 1 & 1 & 1 & 1 \\
Legal retry dispatches & 4 & 0 & 4 & 3 \\
Forbidden repeated requests actually executed & 0/11 & 7/7 & 11/11 & 10/10 \\
\bottomrule
\end{tabularx}
\end{table*}
\begin{table*}[!htbp]
\centering\small
\setstretch{1}
\setlength{\tabcolsep}{5pt}
\renewcommand{\arraystretch}{1.12}
\caption{Additional endpoints for the controlled comparison. All methods use the same 14 completion-designated conditions and three stopping probes. Physical attainment in stopping probes and the all-condition aggregate are descriptive; they do not measure legal completion. Request-level denominators depend on the requests received in each method's trajectory. \texttt{FAIL} denotes an algorithm-defined terminal outcome; implementation errors are counted separately.}
\label{tab:latest-controlled-extra}
\begin{tabularx}{\textwidth}{@{}>{\raggedright\arraybackslash}X rrrr@{}}
\toprule
Metric & ADF-EA & ToolGate & VTC & Direct \\
\midrule
\multicolumn{5}{l}{\textit{Final physical goal met; no completion declaration required}} \\
Completion-designated conditions & 14/14 & 10/14 & 14/14 & 13/14 \\
Stopping probes (descriptive) & 1/3 & 1/3 & 2/3 & 2/3 \\
All conditions (descriptive aggregate) & 15/17 & 11/17 & 16/17 & 15/17 \\
\midrule
Violation-free completion (permitted conditions) & 14/14 & 3/14 & 4/14 & 4/14 \\
Violating physical requests actually executed & 0/14 & 10/10 & 14/14 & 14/14 \\
Legal explicit observation requests incorrectly blocked & 0/7 & 0/4 & 0/5 & 0/5 \\
Safe stop with no earlier violation (three probes) & 3/3 & 2/3 & 0/3 & 0/3 \\
Explicit stop & 3 & 0 & 0 & 0 \\
Algorithm FAIL & 0 & 7 & 0 & 0 \\
Budget cutoff & 0 & 0 & 0 & 0 \\
All repeated dispatches, including legal recovery & 5 & 7 & 13 & 11 \\
Request/dispatch mismatch & 0 & 0 & 0 & 0 \\
Execution errors & 0 & 0 & 0 & 0 \\
\bottomrule
\end{tabularx}
\vspace{4pt}
\begin{minipage}{\textwidth}\footnotesize
\textit{Physical attainment versus authorized execution.} In the budget-reset probe, three dispatches are needed to reach the physical goal but the task budget permits only two. ToolGate, VTC, and Direct reach that goal by making a third, budget-violating dispatch; ADF-EA withholds it. This accounts for VTC's 16/17 versus ADF-EA's 15/17 in the aggregate; both attain 14/14 physical goals in the completion-designated group.
\end{minipage}
\end{table*}
\begin{table*}[!htbp]
\centering\small
\setstretch{1}
\setlength{\tabcolsep}{5pt}
\renewcommand{\arraystretch}{1.12}
\caption{Observed calls and request counts for the controlled comparison, summed over 17 conditions per method. Provider calls and native Boolean actions coincide here. Tokens and continuous-control steps are not applicable.}
\label{tab:latest-controlled-costs}
\begin{tabularx}{\textwidth}{@{}>{\raggedright\arraybackslash}X rrrr@{}}
\toprule
Metric & ADF-EA & ToolGate & VTC & Direct \\
\midrule
Physical-action requests & 67 & 52 & 63 & 63 \\
Provider calls / native Boolean actions & 53 & 52 & 65 & 63 \\
Observations (including algorithm-internal calls) & 7 & 4 & 8 & 5 \\
State reads & 199 & 38 & 22 & 22 \\
All issued requests & 104 & 69 & 93 & 93 \\
Original requests not issued after termination & 0 & 31 & 7 & 7 \\
Model calls & 0 & 0 & 0 & 0 \\
\bottomrule
\end{tabularx}
\end{table*}

ADF-EA enforces different constraints in the three stopping probes. In the budget-reset probe, it rejects the third dispatch because the original two-dispatch budget remains binding, leaving the physical goal unmet. In the alias/retry probe, an authorized retry establishes the processing effect; subsequent revision and redundant-action requests are rejected. The finishing action remains authorized but is absent from the fixed request stream, so the physical goal is unmet when the stream stops. In the obligation-omission probe, the physical goal is met, but missing evidence leaves an earlier obligation unresolved; ADF-EA rejects the completion claim.

The terminal-stop metric requires a stopping endpoint without an accepted completion claim whose physical goal is unmet. The stricter safe-stop metric additionally excludes earlier violating dispatches and unauthorized completion claims. Neither establishes that all legal continuations have been exhausted. The three stopping probes and fourteen completion-designated conditions are predefined groups shared by all methods.

The fixed library contains preparation, processing, and finishing capabilities for each of two state branches, plus a processing-provider alias: seven capabilities in total. ToolGate retains its symbolic state, precondition check, invocation, postcondition check, verified-state update, failed-tool set, and terminal branches. VTC retains the published response branches, tri-valued verification, backoff, bounded retry with $N=1$, and reuse of the idempotency key within an invocation. The shared device has no server-side idempotency service. The only added adapter work is simulator I/O and mechanical binding of declared Boolean preconditions/effects to the algorithms' input functions. No metadata-provenance policy, unresolved-obligation gate, task-budget enforcement, repair scheduler, or task-completion gate is added to either baseline. Complete task and contract inputs remain available to every method. The artifact includes source hashes, input bindings, and the complete 68-run event records.

The reported ADF arm uses the original lifecycle plus a retry hook that consumes only runtime-declared retry authority. It preserves the original stream, including requests that the runtime must reject, and can supply an authorized attempt when the original stream contains no retry request. Rejection alone does not force the loop to terminate. All 68 formal runs use this fixed ADF version and frozen baseline code. Formal and pilot records are independently scored and stored separately.

The composition study uses the runtime version frozen before the retry hook in the controlled comparison. Contracts, mapping, and runtime code have zero per-case edits. F1 executes two prepared branches; F2 combines pre-existing dependencies asymmetrically; F3 interleaves a ready conjunctive node with a fresh chain; F4 adds a false-valued preservation goal; F5 crosses the initial support of two branches; and F6 preserves one completed branch while executing another. F1, F2, F3, F5, and F6 complete for all four methods. F4 fails in ADF during mapping initialization, while the other three methods complete. Its task is feasible; the missing support is the mapping of an already satisfied false-valued goal, not a failure of the candidate action sequence. F4 is included in the denominator, giving ADF-EA 5/6 under the unchanged mapping and runtime. Each method also passes six previously used development cases; these are regression checks, excluded from the new-composition denominator.

The experiment directory retains \texttt{input.json}, \texttt{freeze.json}, \texttt{run\_aligned.py}, \texttt{metrics.py}, and \texttt{test\_metrics.py}. Formal outputs are \texttt{results\_full/raw.jsonl}, \texttt{scores.jsonl}, \texttt{legacy\_scores.jsonl}, \texttt{summary.json}, and \texttt{audit.json}; the pilot is stored separately. The score files distinguish evidence-qualified endpoint completion from violation-free completion. The reproduction guide accompanying this manuscript gives commands and links to the frozen contracts, baseline sources, six-composition configuration, and original logs. The independent event scorer applies to the controlled Boolean comparison; the native, LLM, and SDK studies use the evaluators specified in their respective protocols.

\subsection{Experimental Design and Outcome Measures}
\label{sec:eval-design}

The Boolean Device comparison is specified in Appendix~\ref{app:latest-controlled-details}. Supporting studies use four additional environments with complementary roles. Heating and Room-exit are author-built controlled environments. Heating combines actuation, measurement, and recording to distinguish a physical effect from evidence of that effect: a target temperature may have been reached when measurements are stale, and recording may fail after heating has been verified. Room-exit provides an adaptive task to turn off a light, close a door, and engage a lock. These cases test observation recovery, retention of verified work, and legality along a completing or terminating trajectory.

The two public simulation environments extend the evaluation beyond these purpose-built cases. AI2-THOR supplies household tasks across scenes and object instances, including sequences that open a receptacle, place an object, and close it. This supports broader instance coverage for controlled fault tests and adaptive execution. Meta-World changes the capability interface to bounded manipulation skills, testing whether the assurance rules can be applied while internal control remains with the skill provider. Together, the environments support inspection of individual mechanisms, evaluation across task instances, and reuse through different capability interfaces. All four use software simulation. Household and manipulation outcomes use simulator-native state, while Heating and Room-exit use structured process models. Recorded calls are simulated provider dispatches rather than real hardware operations.

Faults are introduced to distinguish decisions that an acknowledgement alone cannot resolve. A suppressed effect leaves the intended state change absent despite a successful response. Stale evidence leaves the effect in place but requires a new observation. An unavailable successor prevents a later required action. Invalid-proposal tests additionally violate declared ordering, arguments, budgets, or provider-binding conditions. The expected response depends on the study's fault and recovery policy: some conditions permit recovery, while others require withholding further execution. We do not assume that every faulted task should be completed. Corresponding fault categories are deliberately exercised across environments to test the same decision rules under different capability semantics. They are not independent samples of naturally occurring deployment failures.

\begin{table*}[!htbp]
\centering\small
\setstretch{1}
\setlength{\tabcolsep}{6pt}
\renewcommand{\arraystretch}{1.12}
\caption{Evaluation map. Adaptive studies use model-selected requests; controlled studies fix requests or evidence schedules. Sample counts are study-specific and are not pooled.}
\label{tab:claim-evidence}
\begin{tabularx}{\textwidth}{@{}>{\raggedright\arraybackslash}p{0.23\textwidth} >{\raggedright\arraybackslash}p{0.18\textwidth} >{\raggedright\arraybackslash}X >{\raggedright\arraybackslash}p{0.13\textwidth}@{}}
\toprule
\textbf{Evaluation focus} & \textbf{Design} & \textbf{Sample composition} & \textbf{Results} \\
\midrule
\textbf{Published-algorithm comparison}\newline Boolean Device
& Controlled requests; independent event scoring
& 17 conditions on 3 task IDs $\times$ 4 methods = 68 runs; 16 pilot runs excluded
& Table~\ref{tab:latest-controlled-outcomes} \\
\addlinespace[7pt]
\textbf{Frozen-contract composition reuse}
& Six new combinations; frozen runtime
& 6 cases $\times$ 4 methods = 24 records; separate six-case development regression
& Section~\ref{sec:frozen-composition-reuse} \\
\addlinespace[7pt]
\textbf{Completion and safe termination}\newline Room-exit
& Adaptive, Flash and GLM MAX
& 6 configurations $\times$ 3 repetitions $\times$ 2 arms $\times$ 2 models = 72 scheduled runs
& Table~\ref{tab:eval-room-exit} \\
\addlinespace[7pt]
\textbf{Permitted recovery}\newline AI2-THOR ReAct
& Adaptive reevaluation
& 15 tasks, each normal and faulted, $\times$ 2 arms = 60 runs
& Table~\ref{tab:eval-adaptive} \\
\addlinespace[7pt]
\textbf{Admission and verification}
& Fixed-proposal replay
& 43 original and 31 replication tasks, each normal and faulted; separate 49-case lifecycle set
& Table~\ref{tab:eval-controlled} \\
\addlinespace[7pt]
\textbf{ADF continuation mechanisms}
& Internal component ablation
& 43-task faulted population; ADF component removals and Direct reference
& Table~\ref{app:tab:ablation} \\
\addlinespace[7pt]
\textbf{Evidence qualification}
& Contract and evidence perturbations
& 12 paired contrasts, 72 executions; 20 Heating-derived evidence schedules
& Tables~\ref{app:tab:semantic-contrasts}, \ref{app:tab:evidence-stress} \\
\addlinespace[7pt]
\textbf{Recovery permissions}\newline Policy controls
& Scripted request sequences
& 11 cases on 1 task $\times$ 2 arms $\times$ 2 interfaces = 44 runs
& Table~\ref{tab:eval-policy-controls} \\
\addlinespace[7pt]
\textbf{Verified progress after later failures}
& Composed faults and domain cases
& 16 composed-fault tasks; 4 Heating cases; Meta-World: 4 tasks $\times$ 2 seeds under applicable conditions
& Tables~\ref{app:tab:heating-results}, \ref{app:tab:composed-faults}, \ref{app:tab:metaworld-results} \\
\addlinespace[7pt]
\textbf{Native ADF execution cases}
& Development pilot and fixed-request cases
& 6 autonomous AI2-THOR episodes; 3 fixed-request household runs; 2 skill runs
& Sections~\ref{sec:native-agent-mechanism-comparison}, \ref{sec:mechanism-evidence}, \ref{sec:cross-domain-skill-reuse} \\
\addlinespace[7pt]
\textbf{Framework reuse}
& Adaptive reevaluation
& 15 faulted tasks $\times$ 5 SDKs $\times$ 2 arms = 150 runs
& Table~\ref{tab:eval-frameworks} \\
\addlinespace[7pt]
\textbf{Model reuse}
& Model proposals and paired replay
& 15 task instances $\times$ 4 models $\times$ 2 arms = 120 executions
& Tables~\ref{app:tab:model}, \ref{app:tab:model-outcomes} \\
\addlinespace[7pt]
\textbf{Interaction counts}
& Trace accounting
& Adaptive runs above
& Table~\ref{tab:eval-adaptive-cost} \\
\bottomrule
\end{tabularx}
\end{table*}

Controlled paired replay gives both arms the same task, initial state, proposed actions, fault schedule, budget, and evaluator. Direct forwards the retained proposal to the provider. ADF-EA applies its admission, verification, and continuation rules. This tests execution decisions while holding the proposal fixed. Adaptive comparisons instead let each arm choose actions, observations, and termination from its own model trajectory. Direct in these comparisons is an adaptive agent, not a fixed script. It can observe, retry, and change actions. The two arms share the task and model configuration, but their sampled actions and runtime feedback can differ. These are comparisons of the resulting agent systems, not matched-proposal estimates of one isolated runtime component.

Completion is reported at two levels. \emph{Goal-confirmed completion} means that completion is declared or authorized and the evaluator finds the physical goal satisfied. It does not by itself require qualified evidence. \emph{Qualified completion} additionally requires the declared evidence conditions and a protocol-valid completion declaration. For ADF-EA, it also requires runtime completion authorization. The adaptive AI2-THOR reevaluations and policy controls report qualified completion. Replay comparisons explicitly label goal-confirmed completion, including cases where Direct reaches the physical goal without repairing stale evidence. \emph{False completion} refers to a declaration or authorization with an unmet physical goal. Completion accepted with the physical goal met but evidence unqualified is reported as a separate violation.

For the controlled Boolean comparison, \emph{violation-free completion} is an additional, stricter measure: a qualified completion with no earlier policy-violating dispatch. A qualified endpoint can coexist with an earlier improper repeat, so these measures are not interchangeable. Its scorer independently reconstructs native state and evidence from raw events; supporting studies use the evaluators specified below.

\emph{Dispatch violations} count calls that violate declared experimental constraints, including attempts to use an unavailable capability. They do not measure physical injury. In the adaptive reevaluations and recovery-policy controls, we distinguish an agent request, a call entering the provider interface, and an action actually forwarded to the native simulator. A provider retry can follow a failed request that never reached Unity, so it need not repeat a native action. Earlier replay and Room-exit studies report their recorded physical-action dispatch counts, including dispatched attempts that fail. Read-only observations and requests rejected before dispatch are counted separately. In the reevaluations, erroneous rejection is measured over legal physical and agent-observation requests. In the scripted recovery-policy controls, the corresponding denominator contains legal physical requests only. These denominators are not interchangeable.

\emph{Terminal outcomes} distinguish completion, explicit agent termination, budget cutoff, protocol failure, and transport interruption. For Room-exit, \emph{safe termination} denotes ending an infeasible run without claiming completion of an unmet goal; it includes model-issued stops and budget-triggered stops. The legality of preceding dispatches is assessed separately. Controlled \emph{prescribed disposition} measures whether the system selects the response specified for that case, such as rejection, observation, or completion. \emph{Oracle agreement} compares the completion decision with goal truth and also counts correct non-completion when the goal remains unmet. It is not a measure of whether every recoverable task was completed.

Table~\ref{tab:claim-evidence} maps the research questions to study designs, sample composition, and result locations. AI2-THOR supplies household task instances for both replay and adaptive evaluation~\cite{kolve2022ai2thorinteractive3denvironment}. Candidate generation, scene qualification, and the separate grounding samples are detailed in Appendix~\ref{app:evaluation-details}. Heating contributes curated process scenarios, while Meta-World supplies skill-based manipulation tasks~\cite{pmlr-v100-yu20a}. The Room-exit configuration and terminal records are documented separately in Appendix~\ref{app:room-exit-models}. Reusing a task across faults, arms, frameworks, or repeated runs does not create independent task samples, and these study populations are not pooled into one success rate.

The ReAct and five-framework reevaluations use a policy-informed Direct baseline. Both arms receive the task, action groundings, public state, effect/evidence feedback, and the same declared recovery permissions. Direct forwards the agent's physical requests to the provider, while ADF-EA additionally enforces admission and completion authority. The policy allows one retry after qualified evidence of effect failure, provided the provider is available and the request budget remains. It forbids repetition after verified success and provider replacement. Each run allows 12 model calls, eight physical requests, and four evidence reacquisitions in total, counting agent-requested and runtime-initiated observations together. The model chooses physical requests and its terminal decision. ADF-EA may also initiate observation repair. These are system comparisons under a common policy, rather than an isolation of enforcement from feedback and automatic observation.

In these reevaluations, normal execution, transient effect suppression with permitted retry, and stale evidence are designated for completion. The suppressed request fails at the provider interface before reaching Unity, and a permitted retry can execute normally. Stale evidence is injected by replaying pre-action facts in the public evidence channel until a new observation is obtained. In the no-continuation condition, a required downstream provider becomes unavailable and no alternative is permitted. We assess that condition by an explicit agent stop with no preceding policy-violating provider dispatch. The completion conditions use qualified completion as defined above. The final physical assessment ignores provider acknowledgements, but shares its evaluator with public task facts. Evidence assessment reuses the runtime qualifier, so it is not an independent validation of qualifier correctness.

The two reevaluations use the same DeepSeek endpoint configuration, with thinking disabled, temperature zero, and at most 1,024 output tokens per call. Requests specify \texttt{deepseek-v4-flash}, while responses identify \texttt{deepseek-flash}. We disclose this identifier difference without treating it as a verified model alias. Appendix~\ref{app:adaptive-reevaluation-configuration} gives the configuration and accounting details. Every scheduled run is included. Earlier development and historical recordings are kept separate from this complete reevaluation batch, as described in Appendix~\ref{app:framework-recovery-configuration}.

\subsection{Execution Authorization in Adaptive Agent Runs}
\label{sec:eval-adaptive}

The room-exit study asks whether a correct final decision also implies an acceptable execution trace. The agent must turn off the light, close the door, and engage the lock. Three scenario configurations permit completion: normal execution, order-sensitive locking, and a stale door observation that can be refreshed. Three require stopping: a persistently suppressed door-closing effect, an unavailable downstream lock, and a door jammed open. Each configuration is repeated three times per arm, giving nine feasible and nine infeasible runs.

We evaluate the same six configurations with DeepSeek V4.1 Flash and GLM-5.3 MAX. Within each model configuration, the arms share the model, initial public prompt, capability schema, parser, and budgets, while sampling their own action trajectories. Both models show the same execution-assurance benefit: ADF-EA preserves all nine feasible completions, eliminates the six false completions observed with Direct, and prevents unavailable and repeated dispatches in valid trajectories (Table~\ref{tab:eval-room-exit}). All 36 Flash runs are valid. The GLM batch has 34 valid runs and two ADF-EA runs interrupted by transport timeouts in the suppressed-effect condition; these interruptions are retained separately in the results.

\begin{table*}[t]
\centering\small
\setlength{\tabcolsep}{5pt}
\renewcommand{\arraystretch}{1.12}
\caption{Adaptive Room-exit results with two model configurations. Each arm schedules 18 runs. Outcome and execution metrics use valid trajectories, with applicable denominators shown; transport-interrupted runs are reported separately. Safe termination follows Section~\ref{sec:eval-design}, and preceding dispatch violations are assessed separately.}
\label{tab:eval-room-exit}
\begin{tabularx}{\textwidth}{@{}>{\raggedright\arraybackslash}X rrrr@{}}
\toprule
& \multicolumn{2}{c}{DeepSeek V4.1 Flash} & \multicolumn{2}{c}{GLM-5.3 MAX} \\
\cmidrule(lr){2-3}\cmidrule(lr){4-5}
\textbf{Outcome} & Direct & ADF-EA & Direct & ADF-EA \\
\midrule
Valid / scheduled runs & 18/18 & 18/18 & 18/18 & 16/18 \\
Transport-interrupted runs & 0 & 0 & 0 & 2 \\
\midrule
Feasible runs: goal-confirmed completion & 9/9 & \textbf{9/9} & 9/9 & \textbf{9/9} \\
Infeasible runs: safe termination & 3/9 & \textbf{9/9} & 3/9 & \textbf{7/7} \\
False completion & 6/18 & \textbf{0/18} & 6/18 & \textbf{0/16} \\
Effect-failure detection & 4/6 & 6/6 & 0/6 & 4/4 \\
Stale-evidence recovery & 3/3 & 3/3 & 3/3 & 3/3 \\
\midrule
Unavailable dispatches & 7 & \textbf{0} & 8 & \textbf{0} \\
Repeated action dispatches & 12 & \textbf{0} & 5 & \textbf{0} \\
Physical-action dispatches & 67 & 45 & 59 & 41 \\
Requests rejected before dispatch & 0 & 58 & 0 & 62 \\
Model calls in valid trajectories & 89 & 118 & 81 & 113 \\
\bottomrule
\end{tabularx}
\end{table*}

The benefit is visible in both task outcomes and the actions leading to them. With each model, Direct falsely declares completion in all three suppressed-effect runs and all three jammed-door runs. When the downstream lock is unavailable, Direct eventually stops but first dispatches seven unavailable calls with Flash and eight with GLM. ADF-EA instead preserves the unmet obligation and enforces action eligibility before dispatch. Flash safely terminates all nine infeasible runs; GLM explicitly stops in all seven valid infeasible runs. Repeated dispatches fall from twelve to zero with Flash and from five to zero with GLM. Thus, the execution gains accompany full retention of feasible completion, including stale-evidence recovery. Appendix~\ref{app:room-exit-models} reports scenario outcomes, terminal sources, and model configuration.

The AI2-THOR ReAct study tests whether autonomous agents can use the permitted recovery paths as well as stop under unavailable continuation. Each arm runs 15 normal tasks and 15 faulted task instances, with five instances per fault condition. Table~\ref{tab:eval-adaptive} separates conditions expected to complete from those expected to stop. Both arms achieve qualified completion in all 25 permitted runs. All five ADF-EA effect-suppressed runs execute a permitted retry, verify its effect, resolve the failed-effect obligation, and obtain completion authorization. All five stale-evidence runs also complete after evidence recovery.

\begin{table}[t]
\centering\small
\setlength{\tabcolsep}{4pt}
\caption{Adaptive ReAct reevaluation results in AI2-THOR, with 30 scheduled runs per arm. Qualified completion requires the physical goal, qualified evidence, and a valid completion declaration. ADF-EA also authorizes completion. Stopping and the legality of preceding dispatches are assessed separately.}
\label{tab:eval-adaptive}
\par\vspace{6pt}
\begin{tabularx}{\columnwidth}{@{}>{\raggedright\arraybackslash}X rr@{}}
\toprule
Outcome & Direct & ADF-EA \\
\midrule
Normal: qualified completion & 15/15 & 15/15 \\
Allowed recovery: qualified completion & 5/5 & 5/5 \\
Stale evidence: qualified completion & 5/5 & 5/5 \\
\midrule
No continuation: explicit agent stop & 5/5 & 5/5 \\
No continuation: runs with violating dispatch $\downarrow$ & 5/5 & 0/5 \\
\midrule
Policy-violating provider dispatches & 10 & 0 \\
Completion claims with unmet physical goal & 0 & 0 \\
Physical goal met, evidence unqualified at claim & 0 & 0 \\
Legal requests incorrectly blocked & 0/89 & 0/86 \\
Protocol, transport, or budget failures & 0/30 & 0/30 \\
\bottomrule
\end{tabularx}
\end{table}

Both agents explicitly stop in all five no-continuation runs. Direct nevertheless makes ten unavailable provider calls before stopping, while ADF-EA blocks such requests before provider dispatch. These calls do not establish unsafe physical execution inside Unity. Neither arm makes an unsupported completion claim, and ADF-EA incorrectly rejects none of the 86 legal physical or observation requests it receives. All 60 scheduled trajectories end without protocol, transport, or model-budget failure. Thus, the observed benefit is stronger enforcement along the execution trace while preserving normal completion and permitted recovery, rather than a higher completion rate. Section~\ref{sec:eval-cost-boundaries} reports the observation and interaction costs.

\subsection{Controlled Authorization and Evidence Recovery}
\label{sec:eval-attribution}

Controlled replay isolates execution decisions by giving the compared methods the same proposed actions. The 49-case lifecycle study spans eight normal cases, four stale-evidence cases, four suppressed effects, seven unavailable-continuation cases, and 26 invalid proposals. ADF-EA rejects all 26 proposals that violate ordering, budgets, grounding, or binding, and selects the required response in all 49 cases, compared with eight for Direct. This measures appropriate completion, observation, rejection, and stopping across the lifecycle; twelve cases permit completion. Table~\ref{tab:eval-controlled} separates the corresponding outcomes.

The original and fresh AI2-THOR populations test whether the same rules transfer across household instances. Normal completion is preserved in 43/43 and 31/31 tasks with identical physical-action counts between arms. Under injected faults, false completions fall from 15 to zero in the original population and from eight to zero in the replication. ADF-EA rejects missing effects, reacquires stale evidence, and blocks unavailable successors according to the condition. The replication therefore extends the execution-assurance benefit beyond the original task population (Table~\ref{tab:eval-controlled}).

\begin{table*}[t]
\centering\small
\setlength{\tabcolsep}{5pt}
\caption{Controlled execution outcomes. Unified lifecycle contains eight normal
and 41 faulted units. Its fault-only denominators exclude the eight normal
units. These fault categories and rates are specific to each selected population.}
\label{tab:eval-controlled}
\par\vspace{6pt}
\begin{tabularx}{\textwidth}{@{}>{\raggedright\arraybackslash}X cc cc cc@{}}
\toprule
& \multicolumn{2}{c}{Original population} &
\multicolumn{2}{c}{Fresh replication} &
\multicolumn{2}{c}{Unified lifecycle} \\
\cmidrule(lr){2-3}\cmidrule(lr){4-5}\cmidrule(lr){6-7}
Outcome & Direct & ADF-EA & Direct & ADF-EA & Direct & ADF-EA \\
\midrule
Normal goal-confirmed completion & 43/43 & 43/43 & 31/31 & 31/31 & 8/8 & 8/8 \\
Faulted false completion & 15/43 & 0/43 & 8/31 & 0/31 & 8/41 & 0/41 \\
Effect rejection & 0/15 & 15/15 & 0/8 & 8/8 & 0/4 & 4/4 \\
Appropriate no-continuation termination & 0/14 & 14/14 & 0/12 & 12/12 & 0/7 & 7/7 \\
Faulted unsafe/unavailable dispatch & 14/43 & 0/43 & 12/31 & 0/31 & 33/41 & 0/41 \\
\bottomrule
\end{tabularx}

\end{table*}

The internal ablations establish the contribution of ADF's two continuation mechanisms. Removing evidence reacquisition loses all fourteen stale-evidence recoveries; removing continuation authorization restores all fourteen unavailable-successor dispatches. Full ADF-EA retains both benefits. Table~\ref{app:tab:ablation} reports these component removals separately from the published baseline comparison.

Contract perturbations test whether the four DCC blocks change execution decisions. Twelve paired contrasts cover invocation, effect, evidence, and interruption semantics across 72 executions. Intact contracts and sham mutations retain all twelve expected distinctions; targeted weakening retains none. The paired design links each lost distinction to the weakened contract semantics.

Evidence stress tests then examine the verifier's response to timing, dropout, identity, source distinctness, replay, and numeric conflict. Of twenty Heating-derived schedules, eight verify directly and five recover through observation. Six correctly remain unverified: three have insufficient evidence, two inadmissible evidence, and one conflicting evidence. The remaining schedule requires an unsupported measurement-correlation field and is not evaluated. Table~\ref{app:tab:evidence-stress} records the expected and observed disposition of every case. Conflict yields \texttt{UNKNOWN}; a wrong-sample requirement failure denotes inadmissible evidence rather than an absent physical effect (Section~\ref{sec:effect-verification}). These tests show how evidence qualification accepts supported outcomes, repairs recoverable information gaps, and prevents unsuitable evidence from granting execution authority.

\subsection{Policy-Controlled Recovery and Completion}
\label{sec:eval-policy-controls}

The recovery-policy controls test whether one runtime can permit useful recovery and block the same physical request when its policy forbids it. Eleven cases on an AI2-THOR cabinet-opening task in FloorPlan8 cover normal execution, missing evidence, a suppressed first effect with retry allowed or forbidden, provider unavailability, premature completion, and budget limits. Each case is exercised in two arms through the ReAct parser/execution loop and the shared framework tool interface, giving 44 runs: 11 cases $\times$ 2 arms $\times$ 2 interfaces. These controlled request sequences isolate authorization decisions; the installed-SDK adaptive evaluation is reported separately in Section~\ref{sec:framework-portability}.

Both arms receive the same task, policy, initial-state setup, fault rule, observation access, and scripted request sequence. The policy specifies recovery permissions, retry and observation budgets, and completion requirements. Policy-informed Direct exposes this information but forwards requests without ADF-EA's contract enforcement. Thus, the comparison tests enforcement of potentially invalid requests rather than a model's ability to follow policy. Requests continue until the tested interface accepts completion, stops, or reaches its budget. Early acceptance can therefore prevent later scripted requests from being consumed. No online model is called. Appendix~\ref{app:policy-controls} lists the case construction and scoring boundaries.

The key pair holds the missing-effect fault fixed and changes only retry permission. The first provider call acknowledges the request, but fault injection prevents forwarding the command to Unity. Fresh native readback establishes that the cabinet remains closed. With one retry permitted, ADF-EA authorizes the next call, verifies the resulting open state, resolves the obligation, and accepts completion. With retry forbidden, it rejects the same repeated request and leaves the task incomplete. Direct opens the cabinet through the forbidden retry. Its physical success does not make that dispatch compliant with the shared policy.

\begin{table*}[t]
\centering\small
\caption{Recovery-policy controls on one native AI2-THOR task. Counts are per arm across both invocation interfaces. Qualified-completion fractions cover two runs per named case. Forbidden-retry dispatch and incorrect-blocking fractions use request denominators. Remaining counts cover all 22 runs per arm. The forbidden-retry dispatches are included in the total policy violations. Provider calls need not reach Unity.}
\label{tab:eval-policy-controls}
\par\vspace{6pt}
\begin{tabularx}{\textwidth}{@{}>{\raggedright\arraybackslash}X rr@{}}
\toprule
Outcome & Policy-informed Direct & ADF-EA \\
\midrule
\multicolumn{3}{l}{Qualified completion} \\
Normal execution & 2/2 & 2/2 \\
Evidence reacquisition & 2/2 & 2/2 \\
Retry permitted & 2/2 & 2/2 \\
\midrule
Forbidden retry requests dispatched & 2/2 & 0/2 \\
Accepted completion with physical goal unmet & 2 & 0 \\
Accepted completion with goal met but evidence unqualified & 2 & 0 \\
Policy-violating provider dispatches & 6 & 0 \\
Legal physical requests incorrectly blocked & 0/20 & 0/22 \\
Budget-generated cutoffs & 2 & 2 \\
\midrule
Native action executions & 16 & 14 \\
Repeated native action executions & 2 & 0 \\
Explicit observations & 10 & 12 \\
Implicit native reads during method execution & 147 & 282 \\
\bottomrule
\end{tabularx}
\par\smallskip
\parbox{\textwidth}{\footnotesize Legal-request denominators count physical-action requests actually issued and judged admissible under the common policy, not runs or observations. Direct issues 20 and ADF-EA 22. Direct accepts premature completion before any action in two runs, so their later scripted physical requests are not issued. ADF-EA rejects those claims and processes the subsequent legal requests.}
\end{table*}

Table~\ref{tab:eval-policy-controls} separates the three completion-permitted conditions: both arms complete 2/2 normal, 2/2 evidence-reacquisition, and 2/2 retry-permitted runs with qualified evidence. In the paired retry-forbidden case, Direct dispatches both prohibited retry requests and ADF-EA dispatches neither. These two dispatches are part of the six total policy violations prevented by ADF-EA. It also prevents both categories of unsupported completion accepted by Direct. Neither arm incorrectly blocks a legal physical request, with 0/20 for Direct and 0/22 for ADF-EA across all cases. Its two budget cutoffs are reported separately from explicit stopping. All 44 runs finish without protocol errors. These results demonstrate selective execution assurance in the tested controls: enforcement preserves permitted completion while blocking prohibited calls and unsupported completion.

The call trace explains how permitted retry produces the reported result. Fault injection intercepts the first provider request before Unity, so the authorized second request is the first native execution. The table therefore records provider recovery with zero repeated native actions. Scoring uses native state for the physical goal and the common DCC qualifier for evidence, with 282 implicit native reads for ADF-EA versus 147 for Direct. The design demonstrates policy-controlled dispatch and completion at both execution interfaces; autonomous use of the recovery paths is established by the ReAct and SDK studies.

\subsection{Retaining Progress through Failure and Continuation}
\label{sec:eval-continuation}

Composed-fault tasks test whether execution state remains useful when another disruption follows the first. On sixteen native-valid AI2-THOR tasks, ADF-EA selects the prescribed composed outcome in 16/16 versus 0/16 for Direct, with no physical retries. In five cases, an unresolved first effect blocks a later unavailable successor before it is reached. The remaining traces exercise retained progress and successor rejection. This establishes correct decision ordering: earlier unresolved obligations continue to constrain later actions, while still-valid progress remains available.

Heating tests whether recovery can preserve completed work across different device providers. When recording becomes unavailable after heating and measurement, ADF-EA retains the valid prefix and rebinds the unresolved recording step. When heating has occurred but its evidence is incomplete, it gathers measurements without reheating. Across four cases, scenario completion rises from 2/4 to 4/4 and both provider recoveries succeed. The additional work increases physical actions from seven to nine. These scenario-level results demonstrate recovery that completes the remaining task while preserving verified earlier effects; the Heating adapter scores scenarios without a generic terminal \texttt{TaskResult}.

\subsection{Reuse across Models, Agent Frameworks, and Device Domains}
\label{sec:framework-portability}

The grounding study tests how domain knowledge enters the common execution rules. An initial ADF projection rejects order, budget, and availability violations but misses eight target mismatches because the grounding relation is absent. Adding that relation enables ADF-EA to reject 26/26 mismatches and accept 26/26 valid controls on a fresh population, without modifying the assurance core. Tables~\ref{app:tab:grounding-original} and~\ref{app:tab:grounding-refined} retain the original Direct and ADF-EA records. This isolates the adapter's role in supplying physical target meaning to admission and effect verification.

We also vary the upstream model while keeping the execution rules fixed. Qwen3.8 Max, DeepSeek V4 Pro, Kimi K3, and DeepSeek V4.1 Flash each generate valid proposals for all 15 selected task instances. Each retained proposal is then replayed through both Direct and ADF-EA. For every model, ADF-EA preserves all five goal-confirmed completions while reducing false completions from five to zero and runs with violating actions from five to zero. Tables~\ref{app:tab:model} and~\ref{app:tab:model-outcomes} report this consistent execution-assurance benefit across model-generated proposals.

To assess dependence on the upstream framework, we integrate OpenAI Agents SDK~\cite{openai_agents_sdk}, LangGraph~\cite{langgraph}, Google ADK~\cite{google_adk}, Microsoft Agent Framework~\cite{microsoft_agent_framework}, and smolagents~\cite{smolagents}. Each installed SDK runs its own model/tool loop with a custom model adapter connected to the shared decision client. All integrations use the model configuration and explicit recovery policy defined in Section~\ref{sec:eval-design}, the same 15 faulted AI2-THOR tasks, and the same assurance core and contract semantics. Framework-specific prompt and message adaptation remains part of the integration.

We evaluate each framework on three conditions: a missing effect that permits one retry, stale evidence that permits observation repair, and an unavailable downstream provider with no permitted alternative. Each condition contains five task instances per arm, giving 30 runs per framework. Table~\ref{tab:eval-frameworks} reports qualified completion for the two recoverable conditions and the number of runs containing a policy-violating dispatch for the no-continuation condition.

\par
\FloatBarrier
\addvspace{\medskipamount} \noindent\begin{minipage}{\linewidth}
\centering
\small
\setlength{\tabcolsep}{4pt}
\renewcommand{\arraystretch}{1.15}
\captionof{table}{Adaptive results across five agent frameworks. Each condition contains five runs per arm. Qualified completion requires the physical goal, qualified evidence, and a valid completion declaration, with runtime authorization additionally required for ADF-EA. Both arms explicitly stop in every no-continuation run. The final group counts runs containing at least one policy-violating provider dispatch before that stop (lower is better).}
\label{tab:eval-frameworks}
\par\vspace{6pt}
\begin{tabularx}{\textwidth}{@{}>{\raggedright\arraybackslash}X c *{6}{c}@{}}
\toprule
& &
\multicolumn{2}{c}{\shortstack{Allowed retry:\\qualified completion}} &
\multicolumn{2}{c}{\shortstack{Stale evidence:\\qualified completion}} &
\multicolumn{2}{c}{\shortstack{No continuation:\\runs with violating\\dispatch $\downarrow$}} \\
\cmidrule(lr){3-4}
\cmidrule(lr){5-6}
\cmidrule(lr){7-8}
Framework & Version & Direct & ADF-EA & Direct & ADF-EA & Direct & ADF-EA \\
\midrule
OpenAI Agents SDK & 0.19.4 & 5/5 & 5/5 & 5/5 & 5/5 & 5/5 & 0/5 \\
LangGraph & 1.2.9 & 5/5 & 5/5 & 5/5 & 5/5 & 5/5 & 0/5 \\
Google ADK & 2.5.0 & 5/5 & 5/5 & 5/5 & 5/5 & 5/5 & 0/5 \\
Microsoft Agent Framework & 1.12.0 & 5/5 & 5/5 & 5/5 & 5/5 & 5/5 & 0/5 \\
smolagents & 1.26.0 & 5/5 & 5/5 & 5/5 & 5/5 & 5/5 & 0/5 \\
\bottomrule
\end{tabularx}
\end{minipage}
\par\addvspace{\medskipamount}

Across all five frameworks, both arms complete every permitted-retry and stale-evidence run. ADF-EA therefore preserves the tested recovery utility: an authorized retry can establish the missing effect and resolve the outstanding obligation, while observation repair can restore qualified evidence without repeating physical work. These results extend the controlled recovery tests to model-driven execution through five SDK integrations.

In the no-continuation condition, both arms explicitly stop in all five runs per framework. Direct nevertheless calls an unavailable provider in every run, giving 5/5 runs with a violating dispatch and ten such calls per framework. ADF-EA prevents these calls before provider entry, giving 0/5 violating runs. The final columns count runs with a dispatch violation, not failures to stop. Thus, the framework comparison preserves qualified completion in recoverable conditions while reducing violations during runs that must stop.

All 150 scheduled runs finish without protocol, SDK, transport, or budget failure. Neither arm makes a completion claim with an unmet physical goal or unqualified evidence. ADF-EA incorrectly blocks none of the 197 legal physical or observation requests it receives. Together, these results show preserved permitted completion and stronger dispatch enforcement across the tested framework integrations. The shared task instances and model endpoint hold device behavior and model configuration constant as the framework integration changes. Interaction and observation costs are reported separately in Section~\ref{sec:eval-cost-boundaries}, with framework-specific counts in Appendix~\ref{app:framework-details}.

To examine reuse beyond the upstream runtime, we also vary the physical-action domain. Table~\ref{tab:eval-domains} shows how domain-specific contracts, providers, and observation projections connect the same core mechanisms to different meanings of an effect. The domains use intentionally aligned fault categories rather than independently sampled naturally occurring failures.

\begin{table*}[t]
\centering\small
\setlength{\tabcolsep}{4pt}
\caption{Cross-domain mechanism reuse. Different denominators and fault
applicability rules preclude pooling these results into one generalization rate.}
\label{tab:eval-domains}
\par\vspace{6pt}
\begin{tabularx}{\textwidth}{@{}p{0.13\textwidth}*{3}{>{\raggedright\arraybackslash}X}@{}}
\toprule
Domain & Changed integration & Observed mechanism behavior & Scope \\
\midrule
Heating & Process model; heater, measurements, recorder/readback
& Scenario completion 2/4 to 4/4; both applicable provider recoveries succeed
& Four curated software cases; scenario results without generic terminal TaskResult \\
\addlinespace
AI2-THOR & Grounded household actions and native goal predicates
& Controlled replay, held-out replication, and adaptive-agent comparisons
& Qualified tasks under structured observations and injected faults \\
\addlinespace
Meta-World & Bounded scripted skills and native state evaluation
& Normal goal-confirmed completion 8/8 in both arms; all eight missing effects rejected; all six unavailable successors blocked by ADF-EA
& Four tasks, two seeds; no low-level control or real-sensor evaluation \\
\bottomrule
\end{tabularx}

\end{table*}

The domain results show different forms of useful continuation under the same lifecycle. Heating completes provider recovery through two additional actions. Meta-World reacquires qualified evidence in all eight stale cases without a skill retry and blocks every unavailable successor in the six applicable multi-skill plans; two single-skill plans have no successor to test. Across these bindings, contracts specify the effect and evidence conditions while the shared core determines when progress and renewed execution are justified.

These integrations demonstrate reuse of the execution lifecycle through explicit semantic mappings. Contracts and adapters supply domain knowledge, while the core continues to govern admission, qualified effects, progress, and continuation. The framework, model, and domain studies vary their respective factors separately.

\subsection{Task Utility and Interaction Counts}
\label{sec:eval-cost-boundaries}

The interaction study relates execution assurance to useful work and its cost. ReAct preserves all 25 permitted completions per arm, and the five-framework study preserves all fifty. Both agents also stop in every no-continuation run. ADF-EA's advantage is to maintain these task outcomes while preventing the unavailable-provider calls made before Direct stops. Table~\ref{tab:eval-adaptive-cost} distinguishes the model calls, observations, provider requests, and native actions required to obtain those outcomes.

\begin{table*}[t]
\centering\small
\setlength{\tabcolsep}{5pt}
\caption{Interaction and execution counts in the adaptive reevaluations. Columns are totals within each study and arm, not independent task samples. Provider retries are repeated calls entering the provider interface. Native executions count actions forwarded to Unity. Metadata reads count snapshot accesses, not additional physical actions.}
\label{tab:eval-adaptive-cost}
\par\vspace{6pt}
\begin{tabularx}{\textwidth}{@{}>{\raggedright\arraybackslash}X rrrr@{}}
\toprule
& \multicolumn{2}{c}{ReAct: 30 runs per arm} & \multicolumn{2}{c}{Five SDKs: 75 runs per arm} \\
\cmidrule(lr){2-3}\cmidrule(lr){4-5}
Count & Direct & ADF-EA & Direct & ADF-EA \\
\midrule
Model calls & 129 & 122 & 351 & 297 \\
Provider dispatches & 93 & 83 & 245 & 195 \\
Provider retries & 10 & 5 & 50 & 25 \\
Native action executions & 78 & 78 & 170 & 170 \\
Repeated native action executions & 0 & 0 & 0 & 0 \\
Agent-initiated observations & 6 & 3 & 31 & 2 \\
Runtime-initiated observations & 0 & 5 & 0 & 25 \\
Public/runtime metadata reads & 582 & 899 & 1,713 & 3,046 \\
Post-hoc evaluation metadata reads & 55 & 55 & 125 & 125 \\
\bottomrule
\end{tabularx}
\end{table*}

Table~\ref{tab:eval-adaptive-cost} separates the costs of these outcomes. Both arms execute the same number of native actions in each study, with no repeated native execution. The initially suppressed provider request never reaches Unity, so its permitted retry is the first native execution of that attempt's action. Direct's additional provider retries arise in unavailable-continuation conditions and are not additional successful recoveries. ADF-EA performs more metadata reads to support its state and evidence checks. These reads are reported separately from observations that consume the four-reacquisition budget.

ADF-EA reduces model calls from 129 to 122 in ReAct and from 351 to 297 across the five frameworks, decreases provider dispatches, and preserves the same native-action counts. Some observation work moves into the runtime: Direct initiates 37 observations across the two studies, while ADF-EA uses five agent-initiated and thirty runtime-initiated observations. Additional metadata reads support the contract and state checks. These figures measure interaction work; token usage and billing are recorded separately.

Room-exit exposes a different interaction pattern. Flash reduces physical-action dispatches from 67 to 45 while model calls rise from 89 to 118, a 32.6\% increase. GLM uses 81 versus 113 calls in valid trajectories, or 81 versus 131 across all scheduled runs, including the two interrupted ADF-EA trajectories. Its 62 pre-dispatch rejections contain repeated agent proposals without producing disallowed dispatches. The runtime therefore preserves feasible completion and controls device access even when the model requires additional feedback before stopping.

A separate deterministic benchmark tests whether retained progress and late-fault decisions continue to operate as execution depth and evidence width grow. Each dimension varies over \(\{2,4,8,16,32\}\), with five variants per level. Across 100 paired logical units, both arms obtain appropriate normal outcomes in 50/50 cases. On late faults, ADF-EA selects the prescribed outcome in 50/50 versus 0/50 for Direct, demonstrating continued use of earlier progress and evidence obligations through the longer represented executions.

\section{Execution Lifecycle: Implementation Details}
\label{app:execution-lifecycle-details}

This appendix gives the detailed realization of the main-text lifecycle, including proposal authorization, evidence qualification, and continued execution from retained progress. It preserves the implementation-level distinctions and the illustrative Heating trajectory.

\subsection{Authorizing Proposals and Continuation}
\label{sec:authorized-execution}

Materialization resolves the agent candidate against the available capability and task semantics to construct an execution DAG. This exposes ordering, effect dependencies, and evidence requirements for deterministic inspection. Validation checks the represented capability and argument constraints, grounding, ordering, budgets, invocation conditions, and dependencies. A schema-valid candidate can therefore be inadmissible for execution. Dependencies are supplied by or derived from the declared capability and task semantics.

Provider binding determines whether an eligible implementation can realize the validated capability. It considers the provider contract and captured context, including availability, health, adapter readiness, state freshness, and evidence routes. This separates a meaningful semantic request from the availability of an authorized provider. The absence of a binding does not permit substitution outside the boundary. Dispatch follows successful validation and binding. Execution artifacts identify the plan, contract, provider, and occurrence to which the decision applies.

The same boundary applies after feedback. A changed target, parameter, or capability must satisfy the checks for its proposed invocation in the current represented context. The agent's assertion that an earlier action succeeded cannot establish a missing effect. Within supported continuation paths, retained progress can support dependencies under the contract's persistence and verification conditions. Authorization is relative to captured state, so changes after capture require revalidation or provider-level protection.

\subsection{From Qualified Evidence to Verified Progress}
\label{sec:effect-verification}

The verifier establishes whether available observations support the effect of a particular execution. Its inputs are the bound contract and evidence route, execution node and provider, typed observations and gaps, and an evaluation horizon. The route associates a requirement with its predicate, source role, and source provider. Contract fingerprints and execution lineage connect the verification result to the relevant invocation. These associations prevent a successful response or unrelated observation from independently justifying task progress within the supported runtime paths.

Verification has three distinct levels: a requirement conclusion, its interpretation as evidence about an effect, and the resulting execution permission. The shared requirement evaluator returns the conclusion shown below. Its implementation also retains reason codes, selected evidence, and related gaps. These accompanying records are needed to interpret the conclusion.

\[
\begin{aligned}
\operatorname{VerifyReq}(\mathcal C,a,B,t)&\in\mathcal Q,\\
\mathcal Q=\{&\mathrm{SATISFIED},\mathrm{UNSATISFIED},\\
&\mathrm{INSUFFICIENT\_EVIDENCE},\mathrm{UNKNOWN}\}.
\end{aligned}
\]

Here \(a\) is the bound execution context, \(B\) the typed evidence and observation gaps, and \(t\) the evaluation horizon. The procedure first checks requirement--route consistency and the source role's provider-identity relation to the execution provider. Evidence roles requiring provider separation must use a provider identity distinct from the execution provider. Inconsistent artifacts raise an integrity error. Candidate selection then checks route and execution association, freshness, and evidence grade. Some required content matches, including the physical sample identifier in the Heating projection, are evaluated within the requirement predicate rather than removed by this initial selection. The evaluator constructs windows satisfying the required count, identity distinctness, and any consecutive-spacing conditions. Missing eligible observations produce \texttt{INSUFFICIENT\_EVIDENCE}, or \texttt{UNKNOWN} when gaps indicate an unknown state. Window identity uses producer identity when available, otherwise source identity, and finally observation identity. This fallback does not establish distinct physical sensors or statistical independence.

At the requirement level, \texttt{SATISFIED} means that evaluable windows consistently satisfy the requirement, and \texttt{UNSATISFIED} means that they consistently fail it. The failed condition can concern a required identity or field match as well as the intended effect predicate. Conflicting true and false windows produce \texttt{UNKNOWN}. Unevaluable evidence cannot establish an effect. Step-level confirmation requires a nonempty set of requirements that are all satisfied. Otherwise, an unsatisfied requirement takes precedence over unknown and insufficient-evidence conclusions. This aggregation determines whether progress can be promoted. It does not, by itself, classify the physical effect as absent or authorize a retry.

The wrong-sample stress case illustrates this distinction. Its records pass the initial source-route selection, but the required sample-field comparison fails, yielding \texttt{UNSATISFIED} with reason \texttt{FIELD\_MISMATCH}. The experimental harness classifies that result as \emph{inadmissible evidence} for the requested sample. This disposition is not an additional value of \(\mathcal Q\). It means that those measurements cannot establish the requested heating effect, not that qualified measurements show the effect to be absent. Conflicting records sharing one occurrence identity instead trigger an integrity rejection before a requirement conclusion is available. Table~\ref{app:tab:evidence-stress} preserves both the harness dispositions and recorded verifier results.

Verification supports progress promotion only in the associated execution and contract context. Accepted progress can satisfy the dependencies that connect capabilities, while task completion requires all applicable completion obligations. Verification also records what held at an evaluation horizon: a historical fact is not automatically a current prerequisite. Persistence assumptions or fresh evidence determine whether it remains usable. Section~\ref{sec:conditional-soundness} makes this relationship explicit as a theorem of the abstract transition system. Its application to this implementation depends on sound observation interpretation, temporal validity, and faithful mediation; the current experiments do not prove those assumptions for arbitrary devices.

\subsection{Continuing from Progress and Unresolved Obligations}

Repeated action names do not determine recovery authority. A failed attempt with qualified negative evidence may admit a contract retry; an earlier successful state-setting effect that is now false may instead require state repair for the next action. Unknown effects remain unresolved, and still-valid effects do not justify redundant reexecution. The four-way decision in Section~\ref{sec:selective-reexecution} uses retained history together with current evidence. Rule~\ref{rule:state-repair} admits repair through a residual proposal while keeping the original ledger and remaining physical budget; the repair does not spend or replenish the failure-retry allowance. Each successor and the final completion claim still require their own authorization.
\label{sec:continuation-recovery}

Continuation consumes the verification outcome together with interruption rules, current provider eligibility, and bounded recovery authority. When the effect may exist but evidence is insufficient or stale, ADF-EA can authorize additional observation without repeating the physical action. This retains the verification obligation while avoiding the assumption that uncertainty means the action must be executed again. Observation itself remains restricted to the permitted routes and continuation policy.

Physical recovery requires a separate determination that the available evidence supports an unmet intended effect. In the evaluated AI2-THOR recovery adapter, this check supports Boolean equality requirements. It counts observations with the declared source identity and type, target subject, and state path, captured after invocation and within any declared freshness bound. Provider acknowledgements and non-Boolean values are excluded. A relevant requirement supplies qualified negative evidence only when at least the required number of eligible observations contradict its expected Boolean value and none of the eligible observations supports that value. Missing observations, wrong-identity records, stale samples, or conflicting positive and negative readbacks cannot satisfy this check by themselves. This is the implemented check for the evaluated Boolean capabilities, not a generic procedure for negating arbitrary DCC predicates.

Qualified negative evidence still does not grant execution permission. The retry path also requires an unresolved effect obligation, a policy permitting retry, an available provider, satisfied invocation conditions, and remaining retry and action budgets. A permitted retry proceeds through validation, binding, and dispatch. Neither a bare \texttt{UNSATISFIED} label nor rejection of inadmissible evidence bypasses these checks. The implemented direct-rebinding policy separately supports a restricted provider-replacement case under its declared state and budget constraints. If no continuation is admissible, the runtime withholds further dispatch and selects the declared interruption disposition. Actual physical stopping depends on provider behavior. Section~\ref{sec:eval-policy-controls} tests permitted and prohibited retry, while Sections~\ref{sec:eval-adaptive} and~\ref{sec:framework-portability} test agents using the permitted recovery paths.

Table~\ref{tab:method-walkthrough} illustrates how these rules fit together for a heating-and-recording task. It is a schematic walk-through of the supported observation and downstream-recovery mechanisms, not an additional measured trajectory. Heating and recording have separate verification obligations. Successful re-observation can resolve the heating obligation, while a later recording failure leaves the verified heating progress usable if its persistence conditions still hold.

\begin{table*}[t]
\centering\small
\setstretch{1}
\setlength{\tabcolsep}{6pt}
\renewcommand{\arraystretch}{1.15}
\caption{Illustrative execution-state transitions under a contract permitting observation repair and replacement of a failed recording provider. This example explains the method and does not add experimental results.}
\label{tab:method-walkthrough}
\begin{tabularx}{\textwidth}{@{}*{4}{>{\raggedright\arraybackslash}X}@{}}
\toprule
\textbf{Action and observation} & \textbf{Verification result} & \textbf{Progress and obligations} & \textbf{Permitted next step} \\
\midrule
\textbf{1. Heating acknowledged}\newline Authorized heating returns an ACK; temperature evidence is stale.
& Insufficient evidence.
& Heating remains unverified; its evidence obligation stays open.
& Reacquire temperature observations. The ACK does not authorize completion or reheating. \\
\addlinespace[8pt]
\textbf{2. Heating verified}\newline Fresh matching measurements establish the effect.
& Satisfied.
& Heating becomes verified progress; its evidence obligation is resolved.
& Check the recording action's current prerequisites and provider eligibility. \\
\addlinespace[8pt]
\textbf{3. Recording fails}\newline Qualified readback shows the required record is absent.
& Unsatisfied effect requirement, supported by qualified negative evidence.
& The recording obligation stays open; still-valid heating progress is retained.
& Evaluate the declared recovery permission and remaining budget. \\
\addlinespace[8pt]
\textbf{4. Replacement succeeds}\newline An eligible replacement recorder is authorized, executes, and returns qualified readback.
& Satisfied.
& The recording obligation is resolved; both required outcomes are established.
& Authorize completion if the task goal and all required obligations are satisfied. \\
\bottomrule
\end{tabularx}
\end{table*}

Retaining progress does not mean retaining an unconditional permission. If fresh evidence contradicts a required current state, or a contract's freshness or persistence condition no longer holds, the historical verification alone cannot justify the dependent action. That prerequisite must be re-established under the supported continuation rules. This example does not assume automatic progress transfer to an arbitrarily revised task DAG.

Observation repair and physical recovery therefore address different parts of the same execution state. Further observation can discharge a verification obligation without repeating actuation. A physical retry or replacement provider requires a separate admissibility decision. Completion requires the remaining task obligations to be satisfied. These links form the evidence-to-authority lifecycle evaluated below.

\FloatBarrier
\Needspace{8\baselineskip}
\section{Detailed Evaluation Results}
\label{app:evaluation-details}

This appendix reports the task populations and detailed experimental results. The main Evaluation distinguishes controlled replay from independently sampled adaptive runs. Counts in different studies have different applicability rules and are not pooled into a single success rate. In the tables below, goal-confirmed completion for Direct refers to an evaluator-confirmed outcome, not runtime-issued completion authority. Oracle agreement concerns the terminal claim or authority relative to the goal oracle, not the legality of every action.

\FloatBarrier
\Needspace{8\baselineskip}
\subsection{Task populations and controlled execution}

Population construction first qualifies native task feasibility independently of the compared methods. The original study tests 43 tasks under normal and faulted execution; the replication tests 31 new paired tasks. Tables~\ref{app:tab:population}--\ref{app:tab:fresh-results} show how candidate selection leads to those denominators and how normal completion is preserved while false completion and disallowed dispatches are removed under faults. The 49-unit lifecycle set broadens the decision coverage to invalid proposals and condition-specific continuation.

\begin{table*}[!htbp]
\centering
\caption{AI2-THOR population construction. Selection and qualification are performed independently of Direct or ADF-EA execution outcomes.}
\label{app:tab:population}
\par\vspace{6pt}

\small
\begin{tabular}{l r}
\toprule
\textbf{Population stage} & \textbf{Count} \\
\midrule
AI2-THOR scenes & 120 \\
Development scene excluded & 1 \\
Candidate scenes & 119 \\
Stable scenes after repeated reset & 114 \\
Generated task candidates & 25,590 \\
Static eligible candidates & 18,649 \\
Static unresolved candidates & 5,383 \\
Static excluded candidates & 1,558 \\
\midrule
Phase-B stratified sample & 48 \\
Scenes represented in Phase B & 40 \\
Dynamically eligible & 43 \\
Dynamically excluded & 5 \\
Dynamically unresolved & 0 \\
\bottomrule
\end{tabular}
\end{table*}
\FloatBarrier

\begin{table*}[!htbp]
\centering
\caption{Independent AI2-THOR held-out populations used in later evaluations.}
\label{app:tab:heldout-populations}
\par\vspace{6pt}

\small
\begin{tabular}{l r r r}
\toprule
\textbf{Evaluation} &
\textbf{Selected} &
\textbf{Native-valid} &
\textbf{Final paired} \\
\midrule
Fresh execution-time fault replication & 48 & 32 & 31 \\
Unified lifecycle & 64 & 49 & 49 \\
Grounding projection & 32 & 26 & 26 \\
\bottomrule
\end{tabular}
\end{table*}
\FloatBarrier

\begin{table*}[!htbp]
\centering
\caption{AI2-THOR population evaluation under normal and faulted execution.}
\label{app:tab:population-results}
\par\vspace{6pt}

\small
\begin{tabular}{l rr}
\toprule
\textbf{Metric} & \textbf{Direct} & \textbf{ADF-EA} \\
\midrule
\multicolumn{3}{l}{\emph{Normal, $n=43$}} \\
Goal-confirmed completion & 43/43 & 43/43 \\
Oracle agreement & 43/43 & 43/43 \\
Physical actions & 114 & 114 \\
\midrule
\multicolumn{3}{l}{\emph{Faulted, $n=43$}} \\
False completion & 15/43 & 0/43 \\
Goal-confirmed completion & 14/43 & 14/43 \\
Effect rejection & 0/15 & 15/15 \\
Appropriate no-continuation termination & 0/14 & 14/14 \\
Unsafe action & 14/43 & 0/43 \\
Oracle agreement & 28/43 & 43/43 \\
Physical actions & 100 & 86 \\
Observation-only actions & 0 & 14 \\
\bottomrule
\end{tabular}
\end{table*}
\FloatBarrier

\begin{table*}[!htbp]
\centering
\caption{Fresh held-out replication on 31 execution-valid AI2-THOR tasks.}
\label{app:tab:fresh-results}
\par\vspace{6pt}

\small
\begin{tabular}{l rr}
\toprule
\textbf{Metric} & \textbf{Direct} & \textbf{ADF-EA} \\
\midrule
\multicolumn{3}{l}{\emph{Normal}} \\
Goal-confirmed completion & 31/31 & 31/31 \\
Oracle agreement & 31/31 & 31/31 \\
Physical actions & 93 & 93 \\
\midrule
\multicolumn{3}{l}{\emph{Faulted}} \\
False completion & 8/31 & 0/31 \\
Goal-confirmed completion & 11/31 & 11/31 \\
Effect rejection & 0/8 & 8/8 \\
Appropriate no-continuation termination & 0/12 & 12/12 \\
Unsafe action & 12/31 & 0/31 \\
Oracle agreement & 23/31 & 31/31 \\
Physical actions & 82 & 70 \\
Observation-only actions & 0 & 11 \\
\bottomrule
\end{tabular}
\end{table*}
\FloatBarrier

\begin{table*}[!htbp]
\centering
\caption{Unified held-out lifecycle evaluation across 49 native-valid units.}
\label{app:tab:unified-results}
\par\vspace{6pt}

\small
\begin{tabular}{l rr}
\toprule
\textbf{Metric} & \textbf{Direct} & \textbf{ADF-EA} \\
\midrule
Prescribed disposition & 8/49 & \textbf{49/49} \\
Goal-confirmed completion & 24/49 & 12/49 \\
False completion & 8/49 & 0/49 \\
Pre-dispatch rejection & 0/49 & 26/49 \\
Invalid physical dispatch & 26/49 & 0/49 \\
Effect rejection & 0/4 & 4/4 \\
Stale-evidence recovery & 0/4 & 4/4 \\
Appropriate no-continuation termination & 0/7 & 7/7 \\
Unsafe action & 33/49 & 0/49 \\
Oracle agreement & 36/49 & 49/49 \\
Physical actions & 151 & 58 \\
Observation-only actions & 0 & 4 \\
\bottomrule
\end{tabular}
\end{table*}
\FloatBarrier

\FloatBarrier
\Needspace{8\baselineskip}
\subsection{Mechanism controls and semantic coverage}
\label{app:mechanism-controls}

The following internal ADF studies test evidence reacquisition and continuation authorization by removing those components, test domain grounding through the recorded projection variants, and test the four DCC semantic blocks through targeted contract perturbations. They are separate from the public-algorithm baselines. The tables report each study's task set and decision counts.

\FloatBarrier

\FloatBarrier

\begin{table*}[!htbp]
\centering
\caption{Mechanism ablation on the 43-task AI2-THOR faulted population.}
\label{app:tab:ablation}
\par\vspace{6pt}

\small
\begin{tabular}{l rrrr}
\toprule
& \textbf{Direct} &
\textbf{No reacq.} &
\textbf{No cont. auth.} &
\textbf{Full ADF-EA} \\
\midrule
False completion & 15 & 0 & 0 & 0 \\
Goal-confirmed completion & 14 & 0 & 14 & 14 \\
Effect rejection & 0/15 & 15/15 & 15/15 & 15/15 \\
Appropriate no-continuation termination & 0/14 & 14/14 & 0/14 & 14/14 \\
Unsafe action & 14 & 0 & 14 & 0 \\
Unnecessary termination & 0 & 14 & 0 & 0 \\
Oracle agreement & 28/43 & 29/43 & 43/43 & 43/43 \\
Observation-only actions & 0 & 0 & 14 & 14 \\
\bottomrule
\end{tabular}
\end{table*}
\FloatBarrier

\begin{table*}[!htbp]
\centering
\caption{Pre-execution validation of schema-valid corrupted proposals.}
\label{app:tab:grounding-original}
\par\vspace{6pt}

\small
\begin{tabular}{l rr}
\toprule
\textbf{Fault} &
\textbf{Direct invalid dispatch} &
\textbf{ADF-EA reject} \\
\midrule
Order violation & 8/8 & 8/8 \\
Budget violation & 7/7 & 7/7 \\
Unavailable binding & 8/8 & 8/8 \\
Target mismatch, original projection & 8/8 & 0/8 \\
\bottomrule
\end{tabular}
\end{table*}
\FloatBarrier

\begin{table*}[!htbp]
\centering
\caption{Fresh target-grounding evaluation after completing the domain projection.}
\label{app:tab:grounding-refined}
\par\vspace{6pt}

\small
\begin{tabular}{l rr}
\toprule
\textbf{Metric} & \textbf{Direct} &
\textbf{ADF-EA} \\
\midrule
Corrupted proposal rejected & 0/26 & 26/26 \\
Invalid physical dispatch & 26/26 & 0/26 \\
False completion & 22/26 & 0/26 \\
Oracle agreement & 4/26 & 26/26 \\
Corrupted physical actions & 64 & 0 \\
Valid-control acceptance & N/A & 26/26 \\
\bottomrule
\end{tabular}
\end{table*}
\FloatBarrier

\begin{table*}[!htbp]
\centering
\caption{Decision-separation test for the four capability-contract semantics.}
\label{app:tab:semantic-contrasts}
\par\vspace{6pt}

\small
\begin{tabular}{l rrr}
\toprule
\textbf{Semantic dimension} &
\textbf{Intact} &
\textbf{Target weakened} &
\textbf{Sham mutation} \\
\midrule
Invocation & 3/3 & 0/3 & 3/3 \\
Effect & 3/3 & 0/3 & 3/3 \\
Evidence & 3/3 & 0/3 & 3/3 \\
Interruption & 3/3 & 0/3 & 3/3 \\
\midrule
Overall & 12/12 & 0/12 & 12/12 \\
\bottomrule
\end{tabular}
\end{table*}
\FloatBarrier

The 12 paired contrasts yield 72 executions over intact, targeted-weakening, and sham conditions. Their targeted and sham conditions isolate the decision contribution of each semantic block.

\FloatBarrier
\Needspace{8\baselineskip}
\subsection{Cross-domain instantiations}

Heating provides four process-control cases with explicit observation and recovery behavior. Meta-World tests the same lifecycle through bounded manipulation skills. The tables separate domain outcomes from the semantic mappings used to implement them.

\begin{table*}[!htbp]
\centering
\caption{Heating scenarios and the execution-assurance behavior exercised by each case.}
\label{app:tab:heating-cases}
\par\vspace{6pt}

\small
\begin{tabular}{p{1.2cm} p{4.0cm} p{4.3cm} p{5.4cm}}
\toprule
\textbf{Case} & \textbf{Condition} &
\textbf{Direct behavior} & \textbf{ADF-EA behavior} \\
\midrule
H1 &
Normal heating, measurement, and recording &
Completes full proposal &
Requires measurements from a provider distinct from the actuator and record/readback before verified completion
\\

H2 &
\texttt{heater-a} fails with no physical effect &
Stops after failed heater action &
Confirms missing effect, rebinds to \texttt{heater-b}, then completes unresolved suffix
\\

H3 &
Heating effect exists but evidence is insufficient &
Eventually reaches physical completion without typed freshness authority &
Preserves heating and obtains three missing measurements without reheating
\\

H4 &
Heating and measurements complete; recorder unavailable &
Dispatches unavailable recorder and stops &
Preserves verified prefix and rebinds only the unresolved record step
\\
\bottomrule
\end{tabular}
\end{table*}
\FloatBarrier

\begin{table*}[!htbp]
\centering
\caption{Aggregate Heating results. The four cases are curated and are therefore interpreted descriptively.}
\label{app:tab:heating-results}
\par\vspace{6pt}

\small
\begin{tabular}{l rr}
\toprule
\textbf{Metric} & \textbf{Direct} & \textbf{ADF-EA} \\
\midrule
Goal-confirmed completion & 2/4 & 4/4 \\
Effect-failure detection & 0/1 & 1/1 \\
Recovery/rebind & 0/2 & 2/2 \\
Unsafe action & 1/4 & 0/4 \\
Unnecessary termination & 2/4 & 0/4 \\
Physical actions & 7 & 9 \\
Observation-only actions & 11 & 16 \\
\bottomrule
\end{tabular}
\end{table*}
\FloatBarrier

\begin{table*}[!htbp]
\centering
\caption{Meta-World results across four tasks and two seeds.}
\label{app:tab:metaworld-results}
\par\vspace{6pt}

\small
\begin{tabular}{l rr}
\toprule
\textbf{Condition} & \textbf{Direct} & \textbf{ADF-EA} \\
\midrule
Normal goal-confirmed completion & 8/8 & 8/8 \\
Effect-suppressed false completion & 8/8 & 0/8 \\
Effect rejection & 0/8 & 8/8 \\
Explicit stale-observation recovery & 0/8 & 8/8 \\
Physical retry after stale evidence & 0/8 & 0/8 \\
Unavailable successor dispatch & 6/6 & 0/6 \\
Appropriate no-continuation termination & 0/6 & 6/6 \\
\bottomrule
\end{tabular}
\end{table*}
\FloatBarrier

The unavailable-successor condition excludes two single-skill plans as not applicable. Heating uses scenario verification and does not produce the generic terminal \texttt{TaskResult}.

\begin{table*}[!htbp]
\centering
\caption{Realization of the same ADF-EA mechanisms across the three evaluated domains.}
\label{app:tab:domain-mapping}
\par\vspace{6pt}

\small
\renewcommand{\arraystretch}{1.15}
\begin{tabular}{
>{\raggedright\arraybackslash}p{3.0cm}
>{\raggedright\arraybackslash}p{3.5cm}
>{\raggedright\arraybackslash}p{4.2cm}
>{\raggedright\arraybackslash}p{4.2cm}
}
\toprule
\textbf{Mechanism} &
\textbf{Heating} &
\textbf{AI2-THOR} &
\textbf{Meta-World} \\
\midrule

Pre-execution validation / binding &
Detect unavailable recorder and rebind suffix &
Reject order, budget, binding, and grounded-target violations &
Validate and bind retained semantic skill plan
\\[2pt]

Post-action effect verification &
Independently qualify missing heater effect &
Reject suppressed household state transition 15/15 &
Reject missing manipulation effect 8/8
\\[2pt]

Fresh evidence acquisition &
Acquire missing measurements without reheating &
Fresh observation in all stale cases with no physical retry &
Fresh native-state read in 8/8 with no skill retry
\\[2pt]

Continuation / recovery &
Preserve prefix and rebind recoverable suffix &
Terminate when no valid successor remains &
Prevent unavailable successor skill
\\[2pt]

Goal-confirmed completion &
Independent process oracle &
Leaf-wise physical goal and TaskResult &
Native task state and TaskResult
\\[2pt]

\bottomrule
\end{tabular}
\end{table*}
\FloatBarrier

\FloatBarrier
\Needspace{8\baselineskip}
\subsection{Composed faults and evidence perturbations}
\label{app:continuation-details}

The composed-fault study tests how earlier outcomes govern execution after a second disruption; the evidence study tests which observation properties can establish a completed effect. The former reports task-level decision ordering and retained progress. The latter reports direct verification, recovery through observation, or withheld verification under controlled perturbations. Together they connect evidence qualification to the state that governs continued execution.

\begin{table*}[!htbp]
\centering
\caption{Sequential composed-fault evaluation.}
\label{app:tab:composed-faults}
\par\vspace{6pt}

\small
\begin{tabular}{l rr}
\toprule
\textbf{Endpoint} & \textbf{Direct} & \textbf{ADF-EA} \\
\midrule
Correct composed outcome & 0/16 & 16/16 \\
Correct first-fault detection & 0/16 & 16/16 \\
Correct fault precedence & 0/16 & 16/16 \\
Verified-prefix preservation & N/A & 16/16 \\
Unsafe/unavailable dispatch & 11/16 & 0/16 \\
False completion & 5/16 & 0/16 \\
Physical retries & 0 & 0 \\
\bottomrule
\end{tabular}
\end{table*}
\FloatBarrier

The evidence stress study evaluates deterministic decision boundaries using Heating-derived observations, rather than running a simulator or real sensor. The fixed contract requires three identity-distinct matching observations with a one-step cadence, logical age at most 20, and temperature in the inclusive range $[79,81]$ degrees Celsius. Each case assumes a completed heating action. The reported verification therefore concerns whether the supplied evidence justifies that effect, not a newly measured heating success rate. The ADF-EA results cover 20 evidence schedules: eight effects are verified directly and five after additional observation. Table~\ref{app:tab:evidence-stress} reports every schedule, including unverified and unsupported cases.

\begin{table*}[t]
\centering\small
\setstretch{1}
\setlength{\tabcolsep}{6pt}
\renewcommand{\arraystretch}{1.15}
\caption{Case-level evidence stress results for ADF-EA. Expected dispositions come from the predeclared perturbation manifest; observed dispositions and recorded outcomes come from execution records. ``Verify'' concerns heating-effect evidence, not full-task completion.}
\label{app:tab:evidence-stress}
\begin{tabularx}{\textwidth}{@{}>{\raggedright\arraybackslash}X >{\raggedright\arraybackslash}p{0.14\textwidth} >{\raggedright\arraybackslash}p{0.14\textwidth} >{\raggedright\arraybackslash}p{0.18\textwidth}@{}}
\toprule
\textbf{Evidence perturbation} & \textbf{Expected} & \textbf{Observed} & \textbf{Recorded outcome} \\
\midrule
\multicolumn{4}{@{}l}{\textbf{Values and numeric bounds}} \\
\addlinespace[3pt]
Nominal matching measurements & Verify & Verify & Satisfied \\
Measurements spanning both numeric bounds & Verify & Verify & Satisfied \\
Small numeric offsets within the range & Verify & Verify & Satisfied \\
Values exactly at the numeric bounds & Verify & Verify & Satisfied \\
\addlinespace[7pt]
\multicolumn{4}{@{}l}{\textbf{Freshness and cadence}} \\
\addlinespace[3pt]
Logical ages below the limit & Verify & Verify & Satisfied \\
Oldest observation exactly at the age limit & Verify & Verify & Satisfied \\
All observations older than the limit & Observe, verify & Observe, verify & Satisfied \\
Phase shift preserving observation cadence & Verify & Verify & Satisfied \\
Observations exactly at the cadence boundary & Verify & Verify & Satisfied \\
Unequal spacing violates the cadence & Observe, verify & Observe, verify & Satisfied \\
\addlinespace[7pt]
\multicolumn{4}{@{}l}{\textbf{Observation loss}} \\
\addlinespace[3pt]
Temporary loss leaves two observations & Observe, verify & Observe, verify & Satisfied \\
Permanent loss leaves no observations & Insufficient & Insufficient & Insufficient \\
Permanent loss leaves one observation & Insufficient & Insufficient & Insufficient \\
\addlinespace[7pt]
\multicolumn{4}{@{}l}{\textbf{Identity, replay, and lineage}} \\
\addlinespace[3pt]
Correct values with the wrong sample identity & Inadmissible & Inadmissible & Unsatisfied \\
Wrong measurement-correlation identity & Unsupported & Unsupported & Not evaluated \\
Wrong sample and source route & Insufficient & Insufficient & Insufficient \\
Same occurrence replayed three times & Observe, verify & Observe, verify & Satisfied \\
Three records with one producer identity & Observe, verify & Observe, verify & Satisfied \\
Same occurrence with conflicting lineage & Inadmissible & Inadmissible & Integrity rejection \\
\addlinespace[7pt]
\multicolumn{4}{@{}l}{\textbf{Conflicting evidence}} \\
\addlinespace[3pt]
Overlapping strongest windows disagree & Conflict & Conflict & Unknown \\
\bottomrule
\end{tabularx}
\par\vspace{4pt}
\begin{minipage}{\textwidth}
\footnotesize
\textit{Notes.} ``Observe, verify'' denotes verification after observation repair.
The final column reports the recorded requirement conclusion, an integrity rejection,
or non-evaluation; these are distinct from the dispositions in the middle columns.
\end{minipage}
\end{table*}
\FloatBarrier

The six representable but unverified cases match their predeclared dispositions. Permanent total and partial dropout, and the wrong-sample/wrong-route case, retain insufficient evidence. The wrong-sample case fails the matching requirement with \texttt{FIELD\_MISMATCH} and is classified as inadmissible. Conflicting lineage under one occurrence identity produces an integrity rejection, while disagreement between evaluable windows produces \texttt{UNKNOWN}. These outcomes do not supply qualified negative evidence for physical retry (Section~\ref{sec:continuation-recovery}). The measurement-correlation case is not evaluated because its required field is not represented. No physical re-execution is used to resolve uncertainty. The manifest and recorded requirement conclusions make each tested acceptance, observation-repair, and rejection decision inspectable.

\FloatBarrier
\Needspace{8\baselineskip}
\subsection{Recovery-Policy Control Construction and Scoring}
\label{app:policy-controls}

The controls reuse a single F4 state-transition task in FloorPlan8, with the goal of opening a cabinet. Eleven cases specify request sequences for normal completion, a completion claim before any action, repetition after verified success, evidence reacquisition, a premature claim after evidence loss, permitted retry, stopping with retry forbidden, a forbidden retry request, an unavailable provider, exhaustion of the physical-request budget, and exhaustion of the observation budget. The same cases are exercised through the ReAct execution loop and the shared framework tool adapter. Each arm has 22 runs. These are repeated interface checks on one task, not held-out task coverage or statistical repetitions of adaptive behavior.

The common policy permits one effect retry except in the two retry-forbidden cases, which permit none. A retry requires qualified negative evidence and an available provider. Repetition after verified success and provider replacement are prohibited. Evidence reacquisition is allowed within a four-observation budget. The physical-request budget is eight except in the exhaustion case, where it is one. The scripted interaction client uses the existing twelve-turn model-call allowance without making online calls. The permitted and forbidden retry cases suppress the first provider call at the interface, returning an acknowledgement without issuing a native action. Consequently, a later successful provider retry is the first native action execution. The observation-loss cases omit the dispatch-associated evidence sample, while the native physical state remains accessible. Incidental native reads do not automatically repair the missing qualified sample. The explicit observation action supplies the replacement evidence.

Scoring uses the native task evaluator for the physical goal and the shared DCC evidence qualifier for the evidence requirement. The latter is a consistency check against the common contract, not an independent evidence oracle. The controls use structured native state and do not test statistical sensor independence or real-time evidence freshness. Provider-policy violations are assessed against the case's declared retry and availability rules. The recorded false-blocking fraction uses the number of legal physical requests actually issued: 20 for Direct and 22 for ADF-EA. Admissibility is checked from the request-time public state, remaining budget, provider availability, retry limit, and evidence status. These request counts are distinct from the 22 runs per arm. Direct issues two fewer legal physical requests because it accepts completion before any action in the two premature-claim runs. The counts agree with the per-run numbers of provider dispatches after policy-violating dispatches are excluded. Both incorrect-blocking numerators are zero. Native execution counts include only commands forwarded to Unity, and repeated native execution counts additional executions of this task's single physical action. Explicit observations and implicit reads during method execution are reported separately. The two physical-budget cutoffs per arm are not autonomous stopping decisions. Each run retains requests, feedback, provider events, runtime decisions, final scoring, and the associated source and configuration.

\FloatBarrier
\Needspace{8\baselineskip}

\subsection{Adaptive Reevaluation Configuration and Accounting}
\label{app:adaptive-reevaluation-configuration}

The ReAct reevaluation matrix contains the same 15 task identities under normal execution and one assigned fault, giving 60 runs across two arms. The framework matrix applies the 15 faulted task instances to five SDKs, giving 150 runs across two arms. Each fault type has five task instances. The complete batch uses the fixed configuration after development repairs. Development checks and earlier incomplete or failed batches do not contribute rows to these tables. All scheduled runs are retained, without conditioning denominators on successful completion.

Both arms use the official DeepSeek Responses endpoint with requested model \texttt{deepseek-v4-flash}, thinking off, temperature zero, and a 1,024-token output limit. The socket connect/read inactivity timeout is 120 seconds. The full-request size limit is 128,000 UTF-8 bytes, not a token limit. Automatic JSON repair is disabled. Classified transient transport failures permit one additional attempt after two seconds. All 899 actual requests in the complete batch return HTTP 200 with recorded usage, and no transport retry occurs. All responses identify \texttt{deepseek-flash}. The identifier difference is retained as an unverified model-identity mapping, not an official alias confirmation. The batch records 7,086,039 input tokens and 107,762 output tokens. These totals cover both studies and arms and are not a monetary-cost estimate.

The common policy permits one provider retry after qualified negative effect evidence, conditional on provider availability and remaining budget. It does not permit repetition after successful verification or provider replacement. Each trajectory allows 12 model calls, eight physical requests, and four evidence reacquisitions. Agent-requested and runtime-initiated reacquisitions share the latter budget. Incidental metadata accesses are counted separately and are not included in that limit. The stale-evidence injection replays pre-action facts through the evidence channel. It is a controlled observation-repair test, not a test of real-time timestamp expiry or unreliable physical sensing.

Provider dispatches count entries into the provider interface, while native executions count action commands forwarded to Unity. A suppressed provider request can therefore be retried without repeating an action in Unity. Repeated requests rejected before provider entry do not count as provider retries. Rejections are deduplicated by request. A valid completion declaration, evidence qualification, and runtime completion authorization are separate fields. The no-continuation outcome requires an explicit model-issued stop and zero earlier policy-violating provider dispatches. A budget cutoff or interface failure would not satisfy that criterion.

The physical evaluator checks native state without treating provider acknowledgement as proof of an effect. It also supplies public task facts. Evidence scoring reuses the DCC qualifier, so these checks do not independently establish the correctness of the evaluator or qualifier. Model prompts, public feedback, provider calls, observations, authorization outcomes, and terminal decisions are retained for each trajectory. This supports checking the reported execution decisions against the configured policy.

\begin{table*}[!htbp]
\centering\small
\caption{Condition-level interaction counts for the ReAct reevaluation. Each cell is Direct / ADF-EA. Observations are separated by initiator. These are event totals for the indicated number of runs per arm.}
\label{app:tab:react}
\par\vspace{6pt}
\begin{tabularx}{\textwidth}{@{}>{\raggedright\arraybackslash}X c c c c c@{}}
\toprule
Condition & Runs & Model calls & \shortstack{Provider\\dispatches} & \shortstack{Agent\\observations} & \shortstack{Runtime\\observations} \\
\midrule
Normal & 15 & 59 / 59 & 44 / 44 & 0 / 0 & 0 / 0 \\
Allowed effect-failure recovery & 5 & 26 / 27 & 20 / 20 & 1 / 2 & 0 / 0 \\
Stale evidence & 5 & 23 / 18 & 13 / 13 & 5 / 0 & 0 / 5 \\
No legal continuation & 5 & 21 / 18 & 16 / 6 & 0 / 1 & 0 / 0 \\
\bottomrule
\end{tabularx}
\end{table*}
\FloatBarrier

\FloatBarrier
\Needspace{8\baselineskip}
\subsection{Framework Interaction Details}
\label{app:framework-details}

The five installed SDKs own their model/tool loops. Custom model adapters invoke the shared decision client, with framework-specific prompt and message adaptation. The smolagents integration consumes native tool context. The SDKs retain the versions reported in Table~\ref{tab:eval-frameworks}. All 30 scheduled runs per framework finish without a protocol, SDK, transport, or budget failure. This qualification includes real model calls and native AI2-THOR execution, rather than scripted model responses or device fixtures.

\begin{table*}[!htbp]
\centering\small
\caption{Interaction counts through each SDK in the complete reevaluation batch. Each cell is Direct / ADF-EA across 15 runs per arm. Legal requests include physical and agent-observation requests. No legal request is incorrectly rejected in either arm.}
\label{app:tab:framework-functional}
\par\vspace{6pt}
\begin{tabularx}{\textwidth}{@{}>{\raggedright\arraybackslash}X c c c c c@{}}
\toprule
Framework & \shortstack{Model\\calls} & \shortstack{Provider\\dispatches} & \shortstack{Agent\\observations} & \shortstack{Runtime\\observations} & \shortstack{Legal\\requests} \\
\midrule
OpenAI Agents SDK & 70 / 59 & 49 / 39 & 6 / 0 & 0 / 5 & 45 / 39 \\
LangGraph & 70 / 59 & 49 / 39 & 6 / 0 & 0 / 5 & 45 / 39 \\
Google ADK & 70 / 60 & 49 / 39 & 6 / 1 & 0 / 5 & 45 / 40 \\
Microsoft Agent Framework & 70 / 59 & 49 / 39 & 6 / 0 & 0 / 5 & 45 / 39 \\
smolagents & 71 / 60 & 49 / 39 & 7 / 1 & 0 / 5 & 46 / 40 \\
\bottomrule
\end{tabularx}
\end{table*}
\FloatBarrier

\subsection{Execution Assurance across Model-Generated Proposals}
\label{app:cross-model}

The cross-model study uses Qwen3.8 Max, DeepSeek V4 Pro, Kimi K3, and DeepSeek V4.1 Flash on the same 15 AI2-THOR task instances: five with suppressed effects, five with stale evidence, and five without safe continuation. Each model generates a proposal using the declared action transitions. The same retained proposal is then executed through Direct and ADF-EA under the assigned fault. This proposal-then-paired-replay design tests execution assurance across upstream models. All 60 proposals are valid, yielding 60 pairs and 120 arm executions, with no invalid pairs or execution errors.

All four models use OpenClaw with explicit model selection, a requested \texttt{thinking=off} setting, a 300-second timeout, isolated sessions, and no model fallback. Qwen3.8 Max, DeepSeek V4 Pro, and Kimi K3 use the institution-provided Yibu endpoint (\texttt{https://yibuapi.com/v1}) with response identifiers \texttt{qwen3.8-max}, \texttt{deepseek-v4-pro}, and \texttt{kimi-k3}. DeepSeek V4.1 Flash uses the official DeepSeek endpoint with response identifier \texttt{deepseek-flash}. The provider configurations set output-token limits of 8,192 for the Yibu models and 32,768 for Flash. All 236 calls return the requested model and provider identifiers. Table~\ref{app:tab:model} reports proposal generation; Table~\ref{app:tab:model-outcomes} reports the execution outcomes, which are identical across the four models.

\begin{table*}[!htbp]
\centering\small
\setstretch{1}
\setlength{\tabcolsep}{6pt}
\renewcommand{\arraystretch}{1.12}
\caption{Cross-model proposal generation and paired execution. Each model uses the same 15 task instances; each pair contains one Direct and one ADF-EA execution. These are model-generated proposals followed by paired replay.}
\label{app:tab:model}
\begin{tabularx}{\textwidth}{@{}>{\raggedright\arraybackslash}X rrr@{}}
\toprule
\textbf{Model} & \textbf{Valid proposals} & \textbf{Model calls} & \textbf{Execution pairs} \\
\midrule
Qwen3.8 Max & 15/15 & 59 & 15 \\
DeepSeek V4 Pro & 15/15 & 59 & 15 \\
Kimi K3 & 15/15 & 59 & 15 \\
DeepSeek V4.1 Flash & 15/15 & 59 & 15 \\
\bottomrule
\end{tabularx}
\end{table*}

\begin{table}[!htbp]
\centering\small
\setlength{\tabcolsep}{5pt}
\renewcommand{\arraystretch}{1.12}
\caption{Per-model execution outcomes in the cross-model study. Every model in Table~\ref{app:tab:model} produces these same Direct and ADF-EA counts. Denominators are per model, not pooled across models.}
\label{app:tab:model-outcomes}
\begin{tabularx}{\columnwidth}{@{}>{\raggedright\arraybackslash}X rr@{}}
\toprule
\textbf{Outcome} & \textbf{Direct} & \textbf{ADF-EA} \\
\midrule
Goal-confirmed completion & 5/15 & 5/15 \\
False completion & 5/15 & 0/15 \\
Effect-failure rejection & 0/5 & 5/5 \\
Appropriate no-continuation termination & 0/5 & 5/5 \\
Runs with violating actions & 5/15 & 0/15 \\
Oracle agreement & 10/15 & 15/15 \\
Physical-action dispatches & 39 & 34 \\
Observation-only actions & 0 & 5 \\
\bottomrule
\end{tabularx}
\end{table}
\FloatBarrier

For each model, ADF-EA rejects all five suppressed effects, terminates appropriately in all five no-continuation cases, and completes all five stale-evidence cases using observation-only reacquisition. Direct reaches the physical goal in the stale-evidence cases but reports completion for all five suppressed effects and dispatches violating actions in all five no-continuation cases. The results show consistent benefits from contract-directed execution across four upstream models while preserving permitted completion.

\FloatBarrier
\Needspace{8\baselineskip}
\subsection{Room-exit Configuration and Terminal Records}
\label{app:room-exit-models}

The two model configurations each use six deterministic Room-exit scenarios, repeated three times per arm. DeepSeek V4.1 Flash is accessed through OpenClaw 2026.9.2 and the official DeepSeek Chat Completions endpoint, with requested runtime identifier \texttt{deepseek/deepseek-flash} and response identifier \texttt{deepseek-flash}. The adapter requests \texttt{thinking=off}, leaves temperature to the unexposed provider default, and uses a 300-second timeout. Per-run budgets allow 12 model calls, eight physical actions, and four read-only observations. Both arms use the same public prompt, capability schema, and parser. ADF-EA additionally returns public authorization/rejection and capability-unavailable status. Hidden world state, fault identity, and evaluator internals are not exposed to the agent.

For Flash, all 36 trajectories are method-valid. The 207 recorded model calls use isolated sessions and return matching provider and model identifiers, with no model fallback, transport failure, or recorded retry. Automatic JSON repair is disabled. The experiment-local transport permits one retry for classified transient transport failures; none is used. The artifact arm \texttt{REACT\_PLUS\_AUREA} is reported as ReAct + ADF-EA.

GLM-5.3 MAX uses the official Zhipu endpoint \texttt{https://open.bigmodel.cn/api/paas/v4}, runtime identifier \texttt{zhipu/glm-5.3}, and response identifier \texttt{glm-5.3}. The Gateway requests \texttt{thinking=max}; an instrumented outgoing request verifies \texttt{reasoning\_effort=max}, linked to its successful response by session identifier. The task configurations, contracts, prompts, parser, and run budgets remain those above. Each logical model call allows one transport retry after a two-second backoff, with a 300-second timeout per attempt. The two-run pilot is separate from the 36-run study.

The GLM study retains 18 valid Direct trajectories and 16 valid ADF-EA trajectories. Two additional ADF-EA trajectories in the suppressed-door-effect condition end after transport retries are exhausted. The valid ADF-EA trajectories comprise nine goal-confirmed completions and seven model-issued stops, with no budget-triggered termination. Direct records six invalid dispatches and eight unavailable dispatches; ADF-EA records zero of either. Table~\ref{app:tab:room-exit-glm-scenarios} gives scenario-level endpoints.

Across all scheduled GLM runs, 212 logical model calls produce 221 wrapper transport attempts. Eleven attempts time out; nine retries recover seven calls and leave two interrupted trajectories. All 210 successful responses match the requested provider and model, with no fallback or model-protocol parsing error. Returned usage totals 9,409,253 tokens: 569,052 input, 8,095,104 cache-read, and 745,097 output tokens, of which 733,464 are reasoning tokens. Usage missing from timed-out attempts is not counted as zero. Gateway-internal continuations can issue additional requests, so wrapper attempts are not a count of all Provider HTTP requests. Monetary cost is unconfirmed.

\begin{table*}[!htbp]
\centering\small
\setlength{\tabcolsep}{4pt}
\renewcommand{\arraystretch}{1.12}
\caption{Scenario-level DeepSeek V4.1 Flash Room-exit outcomes. The expected endpoint is goal-confirmed completion or safe termination as defined in Section~\ref{sec:eval-design}. Dispatch violations are counted separately.}
\label{app:tab:room-exit-scenarios}
\begin{tabularx}{\textwidth}{@{}>{\raggedright\arraybackslash}X c cc cc cc cc@{}}
\toprule
& & \multicolumn{2}{c}{Expected endpoint} &
\multicolumn{2}{c}{Unavailable calls} & \multicolumn{2}{c}{Repeat dispatches} &
\multicolumn{2}{c}{Action dispatches} \\
\cmidrule(lr){3-4}\cmidrule(lr){5-6}\cmidrule(lr){7-8}\cmidrule(lr){9-10}
Scenario & Expected & Direct & ADF-EA & Direct & ADF-EA & Direct & ADF-EA & Direct & ADF-EA \\
\midrule
Normal & Complete & 3/3 & 3/3 & 0 & 0 & 0 & 0 & 9 & 9 \\
Order-sensitive locking & Complete & 3/3 & 3/3 & 0 & 0 & 0 & 0 & 9 & 9 \\
Suppressed door effect & Stop & 0/3 & 3/3 & 0 & 0 & 3 & 0 & 13 & 6 \\
Stale door evidence & Complete & 3/3 & 3/3 & 0 & 0 & 0 & 0 & 9 & 9 \\
Unavailable downstream lock & Stop & 3/3 & 3/3 & 7 & 0 & 4 & 0 & 13 & 6 \\
Door jammed open & Stop & 0/3 & 3/3 & 0 & 0 & 5 & 0 & 14 & 6 \\
\bottomrule
\end{tabularx}
\end{table*}

\begin{table*}[!htbp]
\centering\small
\setlength{\tabcolsep}{5pt}
\renewcommand{\arraystretch}{1.12}
\caption{Scenario-level GLM-5.3 MAX Room-exit outcomes. Each scenario schedules three runs per arm. Endpoint fractions use valid trajectories; the final column retains the two interrupted ADF-EA runs. Direct's stops in the unavailable-lock condition follow eight unavailable dispatches.}
\label{app:tab:room-exit-glm-scenarios}
\begin{tabularx}{\textwidth}{@{}>{\raggedright\arraybackslash}X c cc c@{}}
\toprule
& & \multicolumn{2}{c}{Expected endpoint} & ADF-EA \\
\cmidrule(lr){3-4}
\textbf{Scenario} & Expected & Direct & ADF-EA & Transport interruptions \\
\midrule
Normal & Complete & 3/3 & 3/3 & 0 \\
Order-sensitive locking & Complete & 3/3 & 3/3 & 0 \\
Suppressed door effect & Stop & 0/3 & 1/1 & 2 \\
Stale door evidence & Complete & 3/3 & 3/3 & 0 \\
Unavailable downstream lock & Stop & 3/3 & 3/3 & 0 \\
Door jammed open & Stop & 0/3 & 3/3 & 0 \\
\bottomrule
\end{tabularx}
\end{table*}

\begin{table*}[!htbp]
\centering\small
\setlength{\tabcolsep}{5pt}
\renewcommand{\arraystretch}{1.12}
\caption{Terminal sources in the two Room-exit studies. The five rows partition each arm's 18 scheduled runs. Model-issued and budget-triggered stops together form the safe-termination endpoint; preceding dispatch violations remain separate.}
\label{app:tab:room-exit-terminals}
\begin{tabularx}{\textwidth}{@{}>{\raggedright\arraybackslash}X rrrr@{}}
\toprule
& \multicolumn{2}{c}{DeepSeek V4.1 Flash} & \multicolumn{2}{c}{GLM-5.3 MAX} \\
\cmidrule(lr){2-3}\cmidrule(lr){4-5}
\textbf{Terminal source} & Direct & ADF-EA & Direct & ADF-EA \\
\midrule
Goal-confirmed completion & 9 & 9 & 9 & 9 \\
False completion & 6 & 0 & 6 & 0 \\
Model-issued stop & 3 & 5 & 3 & 7 \\
Budget-triggered stop & 0 & 4 & 0 & 0 \\
Transport interruption & 0 & 0 & 0 & 2 \\
\bottomrule
\end{tabularx}
\end{table*}
\FloatBarrier

Table~\ref{app:tab:room-exit-terminals} derives terminal sources from the retained step traces. The recorded \texttt{replans} counter counts next-action deviations from the expected sequence (Flash: 8 Direct and 24 ADF-EA; GLM: 6 Direct and 32 ADF-EA), including rejected proposals. In the six valid GLM ADF-EA runs with an unavailable lock or a jammed door, each run rejects ten proposals before the model explicitly stops. These repeated proposals consume interaction budget without causing additional device dispatches. The two transport-interrupted GLM runs contain four additional physical-action dispatches, retained in the trace records separately from the valid-trajectory totals. Physical-action counts refer to simulation dispatch records, including unsuccessful attempts. Repeated dispatches count additional dispatched occurrences of the same action identifier, not requests rejected before dispatch.

\FloatBarrier
\Needspace{8\baselineskip}

\subsection{Execution Outcomes across Depth and Width}

The synthetic benchmark varies execution depth and evidence width to test late-fault decisions and verified-prefix preservation.

\FloatBarrier

\begin{table*}[!htbp]
\centering
\caption{Synthetic execution-depth outcomes and verified-prefix preservation.}
\label{app:tab:scaling}
\par\vspace{6pt}

\small
\begin{tabular}{l rr}
\toprule
\textbf{Endpoint} & \textbf{Direct} & \textbf{ADF-EA} \\
\midrule
Normal appropriate outcome & 50/50 & 50/50 \\
Late-fault appropriate outcome & 0/50 & 50/50 \\
Late-fault false completion & 50/50 & 0/50 \\
Verified-prefix preservation & N/A & 50/50 \\
\bottomrule
\end{tabular}
\end{table*}
\FloatBarrier

\FloatBarrier
\Needspace{8\baselineskip}
\subsection{Additional experimental context}
The initial AI2-THOR census excludes FloorPlan1 as a development scene and retains 114 stable scenes after repeated resets. The Phase-B sample uses 16 family-by-room strata, with three candidates per stratum. Eligibility is determined before Direct--ADF-EA outcomes are observed. The fresh fault replication has one additional initial-state fingerprint mismatch among 32 native-valid cases, leaving 31 paired units without replacement.

\FloatBarrier

The reported counts describe the qualified task populations and the recorded adaptive trajectories. Reusing tasks across fault conditions, frameworks, or repeated runs does not create additional independent task samples. The small curated cross-domain studies and separately varied framework and model studies do not estimate performance over the full deployment population.

\FloatBarrier
\Needspace{8\baselineskip}
\section{Contract Implementation and Assurance Boundaries}
\label{app:implementation-boundaries}

The public DCC schema and provider-bound runtime representation have different roles. Public predicates use the operators \(\{\mathrm{EQ},\mathrm{LT},\mathrm{LTE},\mathrm{GT},\mathrm{GTE}\}\). The public format requires effects and evidence to be nonempty, effect and requirement identifiers to be unique, every effect to have evidence, and interruption causes to be covered exactly once. Parameters and namespaced extensions are optional. Public schema version, provider-contract version, and physical execution identity are distinct. Runtime contracts additionally carry the stable reference, semantic version, lifecycle state, provider/tool identities, input constraints, effect and evidence declarations, and optional restricted direct-rebinding policy.

The shared runtime uses the conflict policy \texttt{STRONGEST\_FRESH\_}\allowbreak \texttt{EVIDENCE\_FAIL\_CLOSED}. For identity-qualified windows, producer identity is used when available, otherwise source identity and finally observation identity. Distinct window identities therefore need not identify distinct physical sensors. Separately, source roles requiring provider separation use different observation and execution provider identities. The implementation names these roles \texttt{INDEPENDENT\_SENSOR} and \texttt{INDEPENDENT\_READ\_BACK}. Neither check establishes statistical independence.

The public ADA-DCC evidence helper has a narrower Boolean interface. It checks source, subject, state path, post-invocation capture, optional age, and predicate satisfaction before counting observations. Its return value must not be equated with the richer runtime's lineage, conflict, and progress decisions. Likewise, the public interruption helper selects a declared disposition. Execution budgets and physical admissibility require runtime enforcement.

Contract semantics and the observation adapters form part of the trusted configuration. The onboarding path requires explicit review decisions and attestations before finalizing a deployable public contract. This checks review completeness, not the reviewer's identity or the physical correctness of the reviewed content. Contract fingerprints detect content inconsistency relative to a captured artifact. They are not signatures establishing a trusted author.

The provider catalog rejects duplicate exact contract identities, identity drift across versions, and multiple active versions of the same reference. Binding excludes contracts whose lifecycle is not \texttt{ACTIVE}, including declared \texttt{EXPIRED} and \texttt{REVOKED} contracts, and checks compatibility against the captured runtime context. These are catalog and binding checks. They do not establish automatic propagation of a later revocation to an already dispatched operation, automatic expiry from wall-clock time, or detection of unreported provider drift. Online updates must not be assumed to revoke existing authority without an explicit revalidation mechanism.

Malformed or inconsistent artifacts can therefore be rejected, and conflicting qualified observations can prevent confirmation. A well-formed but incorrect effect predicate, an undetected stale device limit, or an overly permissive evidence rule can still lead to an incorrect decision. The system does not derive contract truth from contract syntax or from an agent's assertions.

The shared verifier is detailed in Section~\ref{sec:effect-verification}.

The relevant soundness obligation is that a promoted effect corresponds to the intended physical effect. This requires a correct mapping from the task effect to its predicate, trustworthy and correctly associated observations adequate to establish that predicate, and faithful implementation of qualification and promotion. Evidence supporting a historical event also needs appropriate persistence or revalidation conditions before reuse as a current prerequisite. These are assumptions and proof obligations, not an unconditional safety theorem established by schema validation.

Contract coverage can be challenged with paired cases: an effect may be absent after a timeout, or may already have occurred without sufficient evidence. If the declared observations cannot distinguish the cases, the runtime must retain uncertainty and apply an admissible interruption response. Such cases reveal missing semantics but do not prove global decision sufficiency or minimality.

\FloatBarrier
\Needspace{8\baselineskip}
\subsection{Evaluation Versions and Recovery Integration}
\label{app:framework-recovery-configuration}

The ReAct and SDK reevaluation results in this paper come from one complete evaluation after protocol and recovery-integration repairs. Earlier results and development trajectories remain archived separately. The earlier ReAct recording contained seven protocol-error terminations, comprising two normal tasks and five effect-suppressed tasks. In the new configuration, all seven corresponding tasks complete with qualified evidence and ADF-EA authorization, and none reproduces the earlier protocol error. The original seven raw model responses are unavailable, so this is task reevaluation rather than replay of those outputs. The model endpoint, common parser, policy and feedback specification, and recovery integration changed together. The result therefore does not isolate the effect of the parser repair alone.

The earlier five-framework recording also contained effect-suppressed non-completions whose recovery permissions and intervening decisions could not be established from the retained records. The reevaluation explicitly records the common retry policy and its budgets. All 25 effect-suppressed ADF-EA runs across the five SDKs perform permitted recovery and complete. These current results establish the behavior of the revised integration without retroactively classifying the historical terminations as either implementation failures or required stops. No successful development trajectory or selectively rerun task is substituted into the reevaluation result tables.

\end{document}